\documentclass[11pt,DIV=12]{scrartcl}
\usepackage[utf8]{inputenc}
\newif\ifchaos
\newif\ifarxiv
\newif\ifphysicaD
\arxivtrue

\usepackage{microtype}

\ifchaos
\else
\ifphysicaD
\else
    \ifarxiv
    \else
        \usepackage{academicons}
        \usepackage{xcolor}
        \definecolor{orcidlogocol}{HTML}{A6CE39}
    \fi
    \usepackage[misc,geometry]{ifsym}
\fi
\fi

\usepackage[unicode=true,
            bookmarks=true,bookmarksnumbered=false,bookmarksopen=false,
            hidelinks=True,
            colorlinks=false,breaklinks=false,backref=false]{hyperref}

\ifchaos
\else
\ifphysicaD
\else
\hypersetup{pdftitle={Complex pattern formation driven by the interaction of stable fronts in a competition-diffusion system},
            pdfauthor={Lorenzo Contento, Masayasu Mimura},
            pdfstartview=FitV}
\fi
\fi

\usepackage{amsmath}
\usepackage{amssymb}
\usepackage{bm}
\ifchaos
\else
\ifphysicaD
\else
\fi
\fi

\usepackage{amsthm}
\theoremstyle{plain}
\theoremstyle{remark}
\newtheorem{remark}{Remark}

\ifphysicaD
\else
\usepackage{enumerate}
\fi

\usepackage{graphicx}
\graphicspath{{graphics/}}
\ifchaos
\fi
\ifphysicaD
\fi
\ifarxiv
\fi

\usepackage{caption}
\usepackage{subcaption}

\ifchaos
\else
\ifphysicaD
\else
\fi
\fi

\ifchaos
\else
\ifphysicaD
\else
\usepackage{doi}
\usepackage[style=numeric,giveninits=true,uniquename=init,backend=biber,maxcitenames=2,maxbibnames=20]{biblatex}
\AtEveryCite{%
}
\def\cite{\parencite}
\fi
\fi

\ifchaos
\newcommand{\intextcite}[1]{Ref.~\onlinecite{#1}}
\newcommand{\intextcites}[1]{Refs.~\onlinecite{#1}}
\else
\newcommand{\intextcite}[1]{\cite{#1}}
\newcommand{\intextcites}[1]{\cite{#1}}
\fi

\ifchaos
\else
\ifphysicaD
\else
\fi
\fi

\newcommand\shortabstract{%
While it is an established fact that no coexistence
can be observed in the two-species competition-diffusion system
in a convex bounded domain with zero-flux boundary conditions,
under certain hypotheses on the parameter values
the addition of a third species
leads in higher spatial dimensions
to the formation of very complex spatio-temporal patterns
which allow for the competitor-mediated coexistence of all three species.
In order to explain the mechanism behind such patterns,
the interaction of one-dimensional travelling waves is studied,
showing that its outcome (either wave reflection or merging) is deeply linked with the bifurcation structure
of a homoclinic wave and in particular with the appearance of a breathing wave.
Then, the two dimensional complex pattern
can be interpreted as originating from the break-up of a stable regular spiral
due to the reflection of colliding planarly stable fronts.%
}
\newcommand\leadingparagraph{%
Investigation of the mechanisms behind the rich biodiversity
observed in natural ecosystems has always been a central research topic in theoretical ecology.
It is generally thought that species competing for the same limited resources cannot coexist in a homogeneous environment.
This holds true for simple mathematical models of the competition of two species in a convex bounded domain.
However, mathematical ecologists have been recently investigating
whether species coexistence can also occur in a homogeneous environment
thanks to indirect competition dynamics among three or more species.
For example, it is known that,
when a weaker exotic species invades an ecosystem inhabited by two strongly competing native species,
competitor-mediated coexistence can occur
and complex spatio-temporal patterns can be observed in two spatial dimensions.
In this paper we uncover the mechanism which generates such patterns.
First, we show that under some assumptions on the parameters
the three-species Lotka-Volterra competition-diffusion system admits
two planarly stable travelling waves.
We study their interaction in one spatial dimension,
which may result in either reflection or merging into a single homoclinic wave,
depending on the strength of the invading species.
This transition can be understood by studying
the bifurcation structure of the homoclinic wave.
In particular, a time-periodic homoclinic wave (breathing wave)
is born from a Hopf bifurcation
and its unstable branch acts as a separator
between the reflection and merging regimes.
Then, we show how the same transition occurs in two spatial dimensions:
the stable regular spiral associated to the homoclinic wave destabilizes,
giving rise first to an oscillating breathing spiral
and then breaking up producing a dynamic pattern characterized by many spiral cores.
We remark that this process for the generation of complex spatio-temporal patterns
does not require the two interacting fronts to be planarly unstable.%
}
\newcommand\longabstract{%
The ecological invasion problem
in which a weaker exotic species
invades an ecosystem inhabited by two strongly competing native species
is modelled by a three-species competition-diffusion system.
It is known that for a certain range of parameter values
competitor-mediated coexistence occurs
and complex spatio-temporal patterns are observed in two spatial dimensions.
In this paper we uncover the mechanism which generates such patterns.
Under some assumptions on the parameters
the three-species competition-diffusion system admits
two planarly stable travelling waves.
Their interaction in one spatial dimension
may result in either reflection or merging into a single homoclinic wave,
depending on the strength of the invading species.
This transition can be understood by studying
the bifurcation structure of the homoclinic wave.
In particular, a time-periodic homoclinic wave (breathing wave)
is born from a Hopf bifurcation
and its unstable branch acts as a separator
between the reflection and merging regimes.
The same transition occurs in two spatial dimensions:
the stable regular spiral associated to the homoclinic wave destabilizes,
giving rise first to an oscillating breathing spiral
and then breaking up producing a dynamic pattern characterized by many spiral cores.
We find that these complex patterns are generated
by the interaction of two planarly stable travelling waves,
in contrast with many other well known cases of pattern formation
where planar instability plays a central role.%
}

\begin{document}

\ifchaos
\else
\ifphysicaD
\begin{frontmatter}

\title{Complex pattern formation driven by the interaction of stable fronts in a competition-diffusion system}
\author[meiji]{Lorenzo Contento\corref{cor1}}
\ead{lorenzo.contento@gmail.com}
\author[musashino]{Masayasu Mimura}
\cortext[cor1]{Corresponding author}
\address[meiji]{Meiji Institute for Advanced Study of Mathematical Sciences, Meiji University, Tōkyō 164-8525, Japan}
\address[musashino]{Department of Mathematical Engineering, Musashino University, Tōkyō 135-8181, Japan}

\begin{abstract}
\longabstract
\end{abstract}

\begin{keyword}
    competition-diffusion system
    \sep
    ecological invasion
    \sep
    competitor-mediated coexistence
    \sep
    travelling wave
    \sep
    pattern formation
    \sep
    travelling breather
    \MSC[2010] 35Q92 \sep 92D25 \sep 35K57 \sep 35C07 \sep 35B36
\end{keyword}

\end{frontmatter}
\else
    \title{Complex pattern formation driven by the interaction of stable fronts in a competition-diffusion system}
    \author{Lorenzo Contento%
    \thanks{Meiji Institute for Advanced Study of Mathematical Sciences, Meiji University, Tōkyō 164-8525, Japan \newline%
           \Letter\hspace{2mm}\texttt{lorenzo.contento@gmail.com}
           \ifarxiv
           \hspace{3mm}
           \href{https://orcid.org/0000-0002-7901-2172}{ORCID: \texttt{orcid.org/0000-0002-7901-2172}}
           \else
           \hspace{3mm}
           \href{https://orcid.org/0000-0002-7901-2172}{\textcolor{orcidlogocol}{\aiOrcid} \texttt{orcid.org/0000-0002-7901-2172}}
           \fi
    }
    \and Masayasu Mimura
    \thanks{Department of Mathematical Engineering, Musashino University, Tōkyō 135-8181, Japan}
    }
    \maketitle
    \begin{abstract}
    \longabstract
    \medskip
    \newline
    \noindent
    {\scshape Keywords}
    \hspace{0.32mm}
    competition-diffusion system%
    \ifarxiv\ \fi\textperiodcentered\ifarxiv\ \fi
    ecological invasion%
    \ifarxiv\ \fi\textperiodcentered\ifarxiv\ \fi
    competitor-mediated coexistence%
    \ifarxiv\ \fi\textperiodcentered\ifarxiv\ \fi
    travelling wave%
    \ifarxiv\ \fi\textperiodcentered\ifarxiv\ \fi
    pattern formation%
    \ifarxiv\ \fi\textperiodcentered\ifarxiv\ \fi
    travelling breather
    \medskip
    \newline
    \noindent
    {\scshape Mathematics Subject Classification}
    \hspace{0.6mm}
    35Q92%
    \ifarxiv\ \fi\textperiodcentered\ifarxiv\ \fi
    92D25%
    \ifarxiv\ \fi\textperiodcentered\ifarxiv\ \fi
    35K57%
    \ifarxiv\ \fi\textperiodcentered\ifarxiv\ \fi
    35C07%
    \ifarxiv\ \fi\textperiodcentered\ifarxiv\ \fi
    35B36%
    \end{abstract}
    \newpage
\fi
\fi

\section{An introduction to competitor-mediated coexistence in the three-species competition-diffusion system}
\label{sec:introduction}
Investigation of the origin of the rich species biodiversity observed in nature
has been an important and long-standing endeavour in theoretical ecology.
Since Gause's pioneering work on bacterial cultures in the 1930s \cite{gause},
one of the foundations of this field has been the \emph{competitive-exclusion principle}:
two or more species
which are competing for the same limited resources
cannot coexist
if the environmental factors are constant in space and time.
While in many cases species biodiversity in actual ecosystems
is indeed linked to a changing environment or to niche specialization,
there are some examples where a high number of coexisting species
is observed even if resources are scarce and the environmental factors constant
(e.g., Hutchinson's well-known paradox of the plankton \cite{plankton}).
In order to reconcile such an apparent contradiction with the competitive-exclusion principle,
many different mechanisms have been proposed.
Of particular theoretical interest are those that rely only on indirect competition dynamics to enable coexistence,
such as \emph{competitor-mediated coexistence},
where the presence of a third competitor allows two competing species to coexist,
even if they are unable to do so otherwise.

In order to understand competitor-mediated coexistence from a mathematical point of view,
we consider an ecosystem inhabited by three species,
two of which are competing species native to the ecosystem
and the other is an exotic competing species
invading from the outside.
This situation can be described by the following
\emph{three-species competition-diffusion system}
of Lotka-Volterra type
(see, e.g., \intextcites{ikota,morozov}):
\begin{equation}
    \left\{
    \begin{alignedat}{6}
        u_t &= d_1 \, \Delta u &&+ (r_1 &&- u &&- b_{12} \, v &&- b_{13} \, w &&)\,u, \\
        v_t &= d_2 \, \Delta v &&+ (r_2 &&- v &&- b_{21} \, u &&- b_{23} \, w &&)\,v, \\
        w_t &= d_3 \, \Delta w &&+ (r_3 &&- w &&- b_{31} \, u &&- b_{32} \, v &&)\,w,
    \end{alignedat}
    \right.
    \tag{CD}
    \label{eq:cds:pde}
\end{equation}
where $u(x, t)$, $v(x, t)$ and $w(x, t)$ are the population densities
of the three competing species
at the spatial position $x$ and time $t$.
We will take $w$ to be the density of the exotic species
and in the following, with slight abuse of notation,
we will denote by $u,v,w$ both the three competing species and their densities.
The positive parameters $d_i$, $r_i$ and $b_{ij}$,
with $i, j = 1, 2, 3$ and $i \ne j$,
denote the diffusion rates,
the intrinsic growth rates
and the inter-specific competition rates, respectively.
All of them are constant in time and space,
since we want to investigate the effects of indirect competition in a homogeneous environment.
In this paper we are mainly interested in the patterns
displayed by the solutions of the initial value problem \hypertarget{eq:cds:ivp}{(P)}
for the system \eqref{eq:cds:pde}
on a bounded and convex spatial domain $\Omega \subset \mathbb{R}^2$.
The initial conditions are
\begin{equation}
          u(x, 0) = u_0(x),
    \quad v(x, 0) = v_0(x),
    \quad w(x, 0) = w_0(x),
    \quad \text{for all } x\in\Omega,
    \tag{IC}
    \label{eq:cds:ivp:ic}
\end{equation}
where $u_0(x)$, $v_0(x)$ and $w_0(x)$ are given non-negative functions,
and zero-flux boundary conditions are imposed on $\partial\Omega$,
i.e.,
\begin{equation}
      \partial_{\bm{n}} u
    = \partial_{\bm{n}} v
    = \partial_{\bm{n}} w
    = 0,
    \quad
    \text{on } \partial\Omega \times \left(0,\infty\right),
    \tag{BC}
    \label{eq:cds:ivp:bc}
\end{equation}
where $\partial_{\bm{n}}$ denotes the derivative along
the direction normal to the boundary $\partial\Omega$.
Such boundary conditions mean that immigration/emigration is not possible.

In this paper, we consider the invading species $w$
to be ``weaker'' than $u$ and $v$ in their habitat.
This is the ecologically plausible situation
since the native species have evolved there
while the exotic one originated in a different ecosystem.
Thus, we first need to study what it means for a species to be weaker or stronger than another.
We start by considering the relative competitive strength of any couple of species
when they are taken in absence of the third one.
We will use $(u,v)$ as our example, but the same also holds for $(v,w)$ and $(w,u)$.
In absence of the exotic species $w$, i.e., for $w \equiv 0$,
the system \eqref{eq:cds:pde} reduces
to the two-species competition-diffusion system for $u$ and $v$ given by
\begin{equation}
    \left\{
    \begin{alignedat}{5}
        u_t &= d_1 \, \Delta u &&+ (r_1 &&- u &&- b_{12} \, v &&)\,u, \\
        v_t &= d_2 \, \Delta v &&+ (r_2 &&- v &&- b_{21} \, u &&)\,v.
    \end{alignedat}
    \right.
    \label{eq:cds2}
\end{equation}

First, we study what happens if the species are not able to move,
i.e., if we set $d_1 = d_2 = 0$ in \eqref{eq:cds2}.
The resulting system of ordinary differential equations
is the well-known Lotka-Volterra competition system (see, e.g., \intextcite{murray}).
It admits three equilibria:
$(0,0)$, where both species are absent;
$(r_1,0)$, where the first species is alone at its carrying capacity;
$(0,r_2)$, where the second species is alone at its carrying capacity;
and $(u^*,v^*)$, a coexistence equilibrium which is admissible from a modelling point of view only when $u^*,v^* \ge 0$.
The equilibrium $(0,0)$ is always unstable,
while the stability of the others
depends on the parameter values.
Denoting by $\rho = r_2 / r_1$ the relative growth rate of the two species
and by $\tilde{b}_{12} = \rho b_{12}$ and $\tilde{b}_{21} = b_{21} / \rho$
the non-dimensionalized competition rates, four situations can be distinguished:
\begin{enumerate}[(i)]
    \item $\tilde{b}_{12} < 1$ and $\tilde{b}_{21} > 1$.
          Then, $(r_1,0)$ is stable, $(0,r_2)$ is unstable and $(u^*,v^*)$ is not admissible.
          Orbits starting from non-negative initial conditions
          will in general converge to $(r_1,0)$ as time tends to infinity.
          Since the species $u$ always drives $v$ to extinction,
          in this case we say that $u$ is \emph{absolutely stronger} than $v$.
    \item $\tilde{b}_{12} > 1$ and $\tilde{b}_{21} < 1$.
          Then, $(r_1,0)$ is unstable, $(0,r_2)$ is stable and $(u^*,v^*)$ is not admissible.
          Similarly to the previous point, $v$ is absolutely stronger than $u$.
    \item $\tilde{b}_{12} < 1$ and $\tilde{b}_{21} < 1$.
          Then, $(r_1,0)$ and $(0,r_2)$ are unstable, while $(u^*,v^*)$ is admissible and stable.
          Orbits starting from non-negative initial conditions
          will in general converge to $(u^*,v^*)$,
          leading to the coexistence of the species $u$ and $v$.
          Such a case is called \emph{weak competition}.
          At first sight it looks like a violation of the competitive exclusion principle,
          but this situation is generally associated to the existence of extra resources which are not the object of competition.
          For this reason, this case will not be further considered in this paper.
    \item $\tilde{b}_{12} > 1$ and $\tilde{b}_{21} > 1$.
          Then, $(r_1,0)$ and $(0,r_2)$ are stable, while $(u^*,v^*)$ is admissible and unstable.
          Orbits will converge to either $(r_1,0)$ or $(0,r_2)$,
          depending on the particular choice of non-negative initial conditions.
          In this case we say that the species are \emph{strongly competing}.
          Coexistence is impossible but there is no clearly dominant species as in the first two cases.
\end{enumerate}

Now we consider the case in which the species can disperse randomly,
i.e., $d_1,d_2 > 0$ in \eqref{eq:cds2}.
We suppose that the domain $\Omega \subset \mathbb{R}^2$ is convex and bounded
and we impose appropriate initial and boundary conditions
analogous to \eqref{eq:cds:ivp:ic} and \eqref{eq:cds:ivp:bc}.
If $u$ is absolutely stronger than $v$,
then all positive solutions $(u, v)(x, t)$ of the initial value problem
will converge to the spatially-homogeneous equilibrium $(r_1, 0)$.
If $u$ and $v$ are strongly competing instead,
the spatially-homogeneous equilibria $(r_1, 0)$ and $(0, r_2)$ are both stable
and any positive solution generically converges to one of them \cite{hirsch,kishimoto}.
This means that competitive exclusion occurs between the two species $u$ and $v$.
We remark that if $\Omega$ is not convex this is no longer true
and the species may coexist by spatial segregation \cite{matanomimura}.

Given this result, in the case of strong competition
it is natural to wonder
whether it is possible to determine which one of the two species will be dominant in the long run.
As we have already remarked,
in the absence of diffusion
either species may survive,
which one depending on the initial density values.
Thus, in presence of diffusion too,
the surviving species must depend on the initial conditions.
However, let us suppose that at the initial time
the two species are segregated,
in the sense that
the domain can be partitioned in two (not necessarily connected) regions,
in one of which the species $u$ is alone at its carrying capacity
while in the other $v$ is alone at its carrying capacity.
For these quite general and reasonable initial conditions,
the dominant species in the long run can be determined
from the sign of the velocity of the interface between the two regions,
which can be obtained by studying a one-dimensional travelling wave problem
for \eqref{eq:cds2}.
We recall that a travelling wave solution is a solution
whose spatial profile moves at a constant velocity without changing shape.
It has the form $(u,v)(x,t) = (U,V)(x - c \, t)$
for all $x\in\mathbb{R}$ and all $t\in\mathbb{R}$,
where
$c$ is the travelling wave velocity
and $(U,V)(z)$ is the travelling wave profile.
In the case of the one-dimensional travelling waves of \eqref{eq:cds2},
the profile and the velocity
must satisfy
\begin{equation}
    \left\{
    \begin{alignedat}{8}
        &d_1 \, U_{zz} &&+ c \, U_z &&+ (r_1 &&- U &&- b_{12} \, V &&)\,U &&= 0, \\
        &d_2 \, V_{zz} &&+ c \, V_z &&+ (r_2 &&- V &&- b_{21} \, U &&)\,V &&= 0,
    \end{alignedat}
    \right.
    \label{eq:tw2}
\end{equation}
on the whole real line. This system of elliptic equations
is also known as the \emph{travelling wave equation}.
Usually, additional boundary conditions are imposed at $z = \pm \infty$
in order to specify the asymptotic behaviour of the wave far from the front region.
Since we are interested in the motion of the interface
between the region where $u$ is dominant and the region where $v$ is dominant,
we will look for travelling waves of \eqref{eq:cds2}
satisfying the asymptotic boundary conditions
\begin{equation}
    \begin{aligned}
         \lim_{z\to-\infty} (U, V)(z) &= (r_1, 0), \\
         \lim_{z\to+\infty} (U, V)(z) &= (0, r_2).
    \end{aligned}
    \label{eq:tw2_bc}
\end{equation}

In the case of strong competition,
there exists a non-negative and bounded travelling wave solution of \eqref{eq:cds2}
with boundary conditions \eqref{eq:tw2_bc}
and velocity $c = c_{uv}$
which is unique (up to translations) and stable \cite{kanonExistence,kanonStability}.
Then, the sign of $c_{uv}$
determines the direction of the motion of the interface between the species $u$ and $v$.
If for example $c_{uv} > 0$, then the interface moves to the right,
resulting in the species $v$ being completely replaced by $u$
everywhere as $t\to\infty$ in the one-dimensional travelling wave solution.
In such a case we expect that,
for initial conditions where the two species are segregated
and even in higher spatial dimensions,
the species $u$ will always be the one to survive in the long run.
For this reason, if $c_{uv} > 0$ we say that $u$ is \emph{stronger in space} than $v$.
Conversely, if $c_{uv} < 0$, then $v$ is stronger in space than $u$.

\begin{remark}
    In this paper the order of the subscripts in a travelling wave velocity
    reflects the boundary conditions at the spatial infinities.
    For example, in contrast with the symbol $c_{uv}$ introduced in the above paragraph,
    the symbol $c_{vu}$ denotes the velocity of a wave
    whose profile tends to the state where $v$ is alone at carrying capacity as $z \to -\infty$
    and to the state where $u$ is alone at carrying capacity as $z \to \infty$.
    By the uniqueness property discussed above,
    the profile of this wave coincides with that of the wave with velocity $c_{uv}$
    (up to a reflection with respect to the origin and a translation)
    and we have that $c_{vu} = -c_{uv}$.
\end{remark}

If the species $u$ is absolutely stronger than $v$ instead,
intuitively we expect that $u$ is also stronger in space.
In this case, the equilibrium $(0, r_2)$,
which appears in the boundary conditions \eqref{eq:tw2_bc},
is unstable.
Then, it seems reasonable that the interface always moves
in the direction from the stable equilibrium to the unstable one,
since at any point in space the solution tends to the stable state.
In \intextcite{kanonFisher} it was shown that
problem \eqref{eq:tw2} and \eqref{eq:tw2_bc}
admits a continuum of solutions,
meaning that the travelling wave is not unique.
In particular,
there exists a minimal propagation speed $\bar{c} > 0$
such that
problem \eqref{eq:tw2} and \eqref{eq:tw2_bc} admits a solution
for any choice of $c \ge \bar{c}$.
This situation is analogous to the Fisher-KPP equation,
which is also monostable.
As a consequence, all travelling waves have positive velocity
so that the species $u$ always drives the species $v$ to extinction
and $u$ is stronger in space than $v$.
Similarly, if $v$ is absolutely stronger than $u$
then it is also stronger in space.

In the rest of this paper,
we will make the following assumptions on the two native species $u$ and $v$:
\begin{enumerate}[(i)]
    \item $u$ and $v$ are strongly competing, i.e.,
          \begin{equation*} 
              \frac{r_2}{r_1}\,b_{12} > 1
              \quad\text{and}\quad
              \frac{r_1}{r_2}\,b_{21} > 1;
              \tag{A1}
              \label{eq:A1_uv_bi}
          \end{equation*}
    \item $u$ is stronger in space than $v$, i.e.,
          \begin{equation*}
              c_{uv} > 0.
              \tag{A2}
              \label{eq:A2_sign_thetauv}
          \end{equation*}
\end{enumerate}
For example,
if $d_1 = d_2 $, $r_1 = r_2$ and $1 < b_{12} < b_{21}$ are taken,
then clearly both \eqref{eq:A1_uv_bi} and \eqref{eq:A2_sign_thetauv} hold.

Going back to the three-species cases,
we take $r_3$ as a free parameter
and investigate whether competitor-mediated coexistence occurs
when the exotic species $w$ invades the ecosystem inhabited by the native species $u$ and $v$.
Since $r_3$ is the intrinsic growth rate of $w$,
its value reflects the suitability of the new environment for the invader $w$
and thus its strength relative to the native species.
Then, this parameter can be expected
to play a fundamental role in the dynamics after the invasion.

First, we consider two extreme situations for the value of $r_3$.
When $r_3$ is sufficiently large compared with the other parameters,
the exotic species $w$ is absolutely stronger than both native species $u$ and $v$.
In particular, in the ordinary differential equation
obtained in absence of diffusion
the equilibrium $(0, 0, r_3)$ is stable,
while the equilibria $(r_1, 0, 0)$ and $(0, r_2, 0)$ are unstable.
In this case numerical simulations show that
any positive solution $(u, v, w)(t)$ of \hyperlink{eq:cds:ivp}{(P)}
tends to $(0, 0, r_3)$,
which means that the exotic species is always able to invade the new ecosystem,
driving the native species to extinction.
On the other hand, when $r_3$ is sufficiently small,
the invader $w$ is absolutely weaker than both $u$ and $v$
and the equilibria $(r_1, 0, 0)$ and $(0, r_2, 0)$ are stable,
while $(0, 0, r_3)$ is unstable.
In this case, the numerical solutions tend to either $(r_1, 0, 0)$ or $(0, r_2, 0)$,
that is, the invasion is never successful,
the invading species $w$ dies out
and the situation is reduced to the two-species case.
In our case in particular,
thanks to assumptions \eqref{eq:A1_uv_bi} and \eqref{eq:A2_sign_thetauv},
the final state is $(r_1, 0, 0)$ for general initial conditions.
We remark that in the limiting situations where $r_3$ is either very large or very small,
the same results can also be proven analytically
\cite{chm2018}.

Since competitor-mediated coexistence
never occurs for extreme values of $r_3$,
we should consider the case where $r_3$ is neither large nor small
and the exotic species $w$ is of comparable strength with respect to at least one of the native species.
A first possibility is to suppose that the invader $w$ is strongly competing with both native species $u$ and $v$.
Then, in absence of diffusion the equilibria $(r_1, 0, 0)$, $(0, r_2, 0)$ and $(0, 0, r_3)$ are all stable.
Moreover, we find that there exists a unique travelling wave solution
connecting $(0, r_2, 0)$ on the left to $(0, 0, r_3)$ on the right,
having velocity $c_{vw}$ and
for which $u \equiv 0$.
Similarly, there exists a unique travelling wave solution
connecting $(0, 0, r_3)$ on the left to $(r_1, 0, 0)$ on the right,
having velocity $c_{wu}$ and
for which $v \equiv 0$.
Then, if the parameters are chosen
so that $c_{uv}, c_{vw}, c_{wu} > 0$,
we say that \eqref{eq:cds:pde} possesses the \emph{cyclic ordering property} on competition in space.
This means that $u$ invades $v$, $v$ invades $w$ and $w$ invades $u$.
Thanks to this rock-paper-scissors-like relationship among $u$, $v$ and $w$,
one can expect that
the three species may be able to coexist,
exhibiting a dynamic structure of rotating spiral patterns in space.
This situation is discussed more precisely in \intextcite{ikota}.
Instead of considering strongly competing species,
each species may be taken to be absolutely stronger than the next,
from which cyclic competition follows immediately \cite{morozov}.

In this paper we are concerned with the ecologically realistic case of
an invading species $w$ weaker than the native species $u$ and $v$.
Thus, the setting described above is not the correct one.
As a more realistic situation,
we assume that $w$ is absolutely weaker than $v$
but strongly competing (not absolutely stronger) with $u$.
The former holds if
\begin{equation*}
    \frac{r_2}{r_3} \, b_{32} > 1 > \frac{r_3}{r_2} \, b_{23},
    \tag{A3}
    \label{eq:A3_vw_mono}
\end{equation*}
while the latter holds if
\begin{equation*}
    \frac{r_1}{r_3} \, b_{31} > 1
    \quad\text{and}\quad
    \frac{r_3}{r_1} \, b_{13} > 1.
    \tag{A4}
    \label{eq:A4_uw_bi}
\end{equation*}
Under \eqref{eq:A1_uv_bi}, \eqref{eq:A3_vw_mono} and \eqref{eq:A4_uw_bi},
in the diffusion-free system associated to \eqref{eq:cds:pde}
the equilibria $(r_1, 0, 0)$ and $(0, r_2, 0)$ are locally stable,
while $(0, 0, r_3)$ is saddle-type unstable.

Since we want to focus on coexistence
originating from the interplay
of competition dynamics and diffusion,
we need to ensure that
coexistence is not possible
in the trivial case
when diffusion is absent.
This is done by imposing the following additional assumption:
\begin{equation}
    \begin{gathered}
        \ifphysicaD
        \text{In the diffusion-free system associated to \eqref{eq:cds:pde},}
        \\
        \text{the positive coexistence equilibrium}
        \\
        \text{either does not exist or is saddle-type unstable.}
        \else
        \text{In the diffusion-free system associated to \eqref{eq:cds:pde},
              the positive coexistence equilibrium}
        \\
        \text{either does not exist or is saddle-type unstable.}
        \fi
    \end{gathered}
    \tag{A5}
    \label{eq:A5_no_internal_eq}
\end{equation}
Then, under assumptions (A1) and (A3--5),
positive solutions $(u, v, w)(t)$
of the diffusion-free system
associated to \eqref{eq:cds:pde}
tend in general to either $(r_1, 0, 0)$ or $(0, r_2, 0)$
\cite{hofbauer,zeeman}.

From this result, it seems reasonable to expect
that competitor-mediated coexistence never occurs
even when the species can move randomly in space.
Indeed, this is the case
if all the diffusion rates $d_i$ ($i = 1, 2, 3$) are very large \cite{conway},
that is, all orbits of \hyperlink{eq:cds:ivp}{(P)} converge
to one of the two spatially-homogeneous equilibria
$(r_1, 0, 0)$ and $(0, r_2, 0)$.
However, we will show that this is no longer the case
if the diffusion rates $d_i$ are sufficiently small
and that it is possible for $u$, $v$ and $w$ to coexist
even if the invader $w$ is weaker, i.e., cannot survive,
in the diffusion-free system.
In order to do that, we set the parameter values in \eqref{eq:cds:pde} as
\begin{equation}
    \begin{aligned}
        & d_1 = d_2 = d_3 = 1, \\
        & r_1 = r_2 = 28,   \\
        & \begin{aligned}
        b_{12} &= 22/21, & b_{13} &= 4,     \\
        b_{21} &= 1.87,  & b_{23} &= 3/4,   \\
        b_{31} &= 26/21, & b_{32} &= 22/21, \\
        \end{aligned}
    \end{aligned}
    \label{eq:parameters}
\end{equation}
and take $r_3$ as a free parameter \cite{tohma,contento}.
It can be shown that,
as long as $r_3 \in \left( 7, 29+1/3\right)$,
the assumptions (A1--5) are satisfied.

\ifchaos
\begin{figure*}
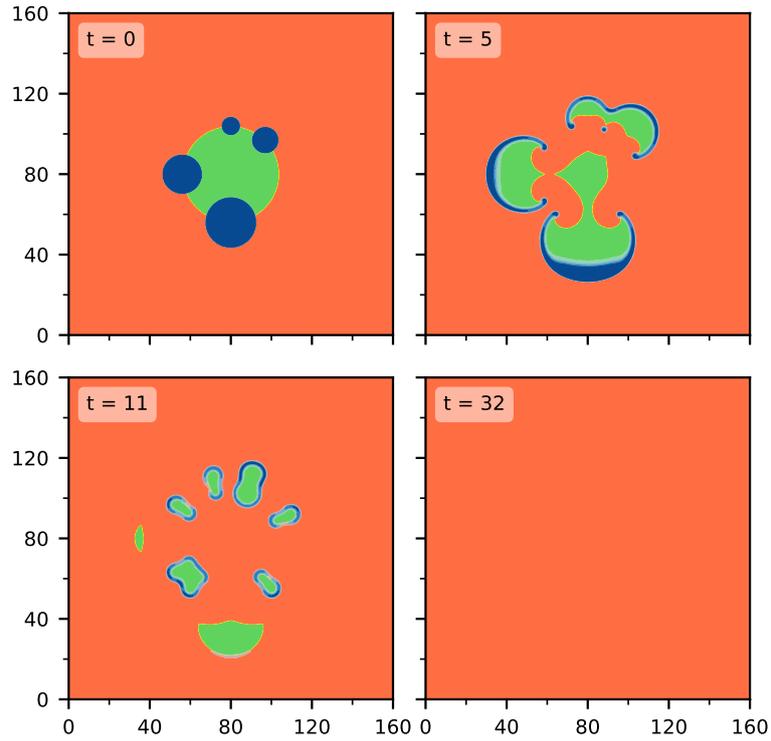

\else
\begin{figure}[p]
\fi
    \centering
    \includegraphics[width=0.6666\textwidth]{{{invasion__r3=26.55_L=160}}}
    \caption{Invasion by the exotic species $w$ when $r_3=26.55$.
             Areas where $u>v$ (respectively, $v>u$)
             are plotted in red (respectively, green)
             and the interface between the two is highlighted in yellow.
             On top of this layer
             $w$ has been superimposed in blue colour,
             the darker the higher its density.
             More details can be observed by zooming in in the electronic version.}
    \label{fig:r_3_small}
\ifchaos
\end{figure*}
\else
\end{figure}
\fi
\ifchaos
\begin{figure*}
\else
\begin{figure}[p]
\fi
    \centering
    \includegraphics[width=\textwidth]{{{invasion__r3=26.75_L=160}}}
    \caption{Invasion by the exotic species $w$ when $r_3=26.75$. Initial conditions are the same as in Figure~\ref{fig:r_3_small}.}
    \label{fig:r_3_mid}
\ifchaos
\end{figure*}
\else
\end{figure}
\fi
\ifchaos
\begin{figure*}
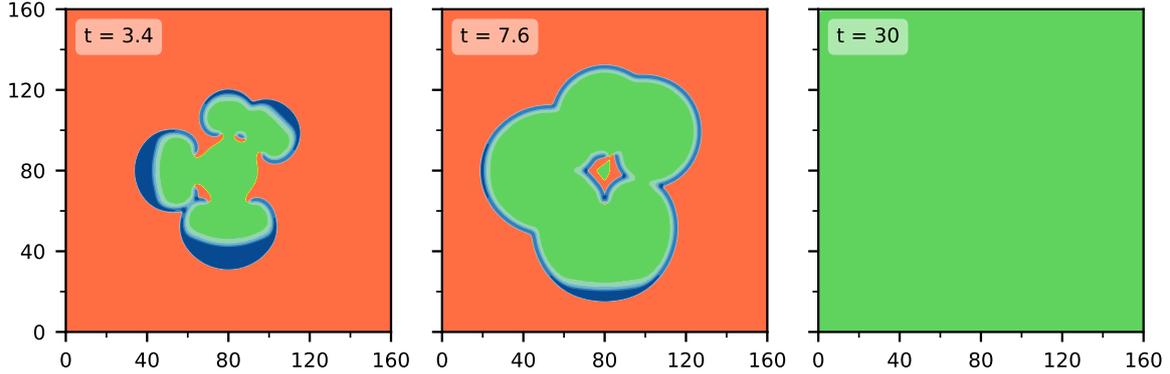

\else
\begin{figure}
\fi
    \centering
    \includegraphics[width=\textwidth]{{{invasion__r3=28.00_L=160}}}
    \caption{Invasion by the exotic species $w$ when $r_3=28$. Initial conditions are the same as in Figure~\ref{fig:r_3_small}.}
    \label{fig:r_3_large}
\ifchaos
\end{figure*}
\else
\end{figure}
\fi

We consider a square domain $\Omega$
and the initial conditions shown in the first panel of Figure~\ref{fig:r_3_small},
where the exotic species $w$ (in blue colour) invades near the interface
between the native species $u$ and $v$ (in red and green colour, respectively).
In absence of $w$ the region inhabited by $v$ would shrink by competitive exclusion
and $u$ would eventually occupy the whole domain $\Omega$.
This is still true even with the addition of $w$
when $r_3$ is relatively small,
as shown in Figure~\ref{fig:r_3_small} for $r_3 = 26.55$.
On the other hand, when $r_3$ is relatively large,
the species $v$ will be the only one surviving at the end,
as can be seen in Figure~\ref{fig:r_3_large} for $r_3 = 28$.
Note that also in this case competitive exclusion occurs,
but the final outcome differs from the one in Figure~\ref{fig:r_3_small}:
thanks to the presence of the invading species $w$ and cyclic competition,
$v$ is able to replace $u$ as the dominant species in the long run.
Finally, when $r_3$ has a suitable intermediate value,
a complex spatio-temporal pattern appears
and competitor-mediated coexistence can be observed,
as shown in Figure~\ref{fig:r_3_mid} for $r_3 = 26.75$.
Animations for the simulations
in Figures~\ref{fig:r_3_small}, \ref{fig:r_3_mid} and \ref{fig:r_3_large}
can be found at \intextcite{movieInvasion}.

From the simulation results shown in Figures~\ref{fig:r_3_small}, \ref{fig:r_3_mid} and \ref{fig:r_3_large},
it is clear that the patterns produced by the the three-species competition-diffusion system \eqref{eq:cds:pde}
depend sensitively on the value of $r_3$.
From a mathematical point of view, the understanding of the nature of this dependence,
and in particular determining the conditions for which competitor-mediated coexistence occurs,
is a problem of great interest.

We should note that
complex spatio-temporal patterns
such as labyrinthine patterns, spot replications and spiral turbulence
have been already observed in other reaction-diffusion systems.
The mechanisms generating such patterns are diverse
and have been intensively studied from both experimental and analytical viewpoints.
Given the wealth of examples in the literature
we will just give a far from exhaustive list
and mention only some of the most relevant results.

As can be seen in Figure~\ref{fig:r_3_mid},
spiral cores appear to be an important feature of the patterns displayed by our three-species competition-diffusion system.
Regular and stable spirals often appear in excitable and oscillatory systems, such as the FitzHugh-Nagumo system,
in correspondence to a stable travelling pulse.
As parameters are varied, such spirals may destabilize
(either near the core or far away from it).
Then, the spiral often breaks up
leading to complex spatio-temporal patterns (\emph{spiral turbulence}, e.g., \cite{zhou,bar,yang})
resembling those in Figure~\ref{fig:r_3_mid}.
More in general, instabilities are often capable of generating complex patterns.
For example, we refer to \cite{nagayama} for the case of an unstable expanding ring in a bistable reaction-diffusion system
and to \cite{reynolds,lee} for spot instabilities in autocatalytic reaction-diffusion systems.

As we will see in the following,
another feature of the patterns shown in Figures~\ref{fig:r_3_small}, \ref{fig:r_3_mid} and \ref{fig:r_3_large}
is that they can be thought as being generated from the interaction of two stable counter-propagating fronts.
Also in bistable FitzHugh-Nagumo-like systems,
depending on the parameter values,
it is possible to observe a similar couple of stable counter-propagating fronts,
which are generated from a pitchfork bifurcation often referred to as the nonequilibrium Ising-Bloch (NIB) bifurcation
(see, e.g., \cite{hagberg1993, hagberg1994}).
These two fronts can form regular spirals and
these spirals can destabilize by planar instability,
again generating complex patterns of various shapes \cite{meron}.
The behaviour of such fronts in response to curvature perturbations
and the consequent spiral nucleation has been studied in \cite{elphick1995, hagberg1997, hagberg1998}.
NIB bifurcations and the associated patterns have also been observed experimentally in several settings,
such as parametrically-forced oscillating systems \cite{coullet1990, coullet1992, elphick1997, marts2004, barashenkov2005, gomila2015},
bistable chemical reactions \cite{haim1996}, liquid crystals \cite{migler1994, frisch1995}
and non-linear optical systems \cite{perez2004, esteban2005, sanchez2005}.

Keeping these earlier studies in mind,
we will shed light on the mechanism behind the complex dynamical pattern of Figure~\ref{fig:r_3_mid}.
Our approach will be to first study the one-dimensional case,
in particular travelling waves and their interaction,
and then use the insight so obtained to explain the two-dimensional case.

\section{One-dimensional travelling wave solutions}
\label{sec:1d_tw}
Since the species $v$ is absolutely stronger than $w$ by \eqref{eq:A3_vw_mono},
we know that $c_{vw} > 0$,
where $c_{vw}$ is the non-unique velocity of the travelling wave solutions
connecting $(0,r_2, 0)$ and $(0, 0, r_3)$ \cite{kanonFisher}.
Since the species $w$ and $u$ are strongly competing by \eqref{eq:A4_uw_bi} instead,
there exists a unique travelling wave solution
connecting $(0, 0, r_3)$ and $(r_1, 0, 0)$
with velocity $c_{wu}$ \cite{kanonExistence,kanonStability}.
We will assume that
\begin{equation*}
    c_{wu} > 0,
    \tag{A6}
    \label{eq:A6_sign_thetawu}
\end{equation*}
which is satisfied for the parameter choices given in \eqref{eq:parameters}.

Assumptions \eqref{eq:A2_sign_thetauv}, \eqref{eq:A3_vw_mono} and \eqref{eq:A6_sign_thetawu}
imply that \eqref{eq:cds:pde} possesses the cyclic ordering property on competition in space.
Then, we may expect species coexistence to occur,
similarly to \intextcites{ikota,morozov} where the same ordering property applies.
This is indeed the case, as we have shown at the end of the previous section.
In particular, by looking closely at Figures~\ref{fig:r_3_mid} and \ref{fig:r_3_large},
we see that $\Omega$ is essentially partitioned in two regions:
one where the species $u$ is dominant and $(u,v,w) \approx (r_1,0,0)$
and the other where $v$ is dominant instead and $(u,v,w) \approx (0,r_2,0)$.
If, somehow arbitrarily, we define these two regions as the ones
where $u > v$ and $v < u$ respectively,
then the interface between them is given by the line where $u = v$.
By examining again the simulation results,
we see that this interface
can be classified into two kinds.
The first one is observed in the absence of the invader $w$
(yellow interfaces in the figures)
and moves in the direction from $u$ towards $v$.
This interface is the two-dimensional analogous
of the one-dimensional travelling wave
connecting $(r_1,0)$ and $(0,r_2)$
in the two-species competition-diffusion system \eqref{eq:cds2}.
On the other hand, the second kind of interface is characterized
by the presence of the invading species $w$ near the front
(blue interfaces in the figures).
Such interfaces move in the direction from $v$ towards $u$
and result in the usually weaker species $v$ displacing the usually stronger species $u$,
reversing the ``natural'' strength relationship occurring in absence of $w$.
This is possible thanks to the cyclic competition in space,
which allows for $w$ to initially replace $u$
and for $v$ to replace $w$ immediately after.
Moreover, it suggests that there exists
a travelling wave solution
connecting $(r_1,0,0)$ and $(0,r_2,0)$
in which all species are present.

Under our assumptions,
equilibria $(r_1, 0, 0)$ and $(0, r_2, 0)$ are stable
in the system of ordinary differential equations
obtained from \eqref{eq:cds:pde} in absence of diffusion.
Now we discuss the existence of travelling wave solutions of \eqref{eq:cds:pde}
connecting these two equilibria.
The profile $(U,V,W)(z)$ and velocity $c$
of such a travelling wave
must satisfy the travelling wave equation
\begin{equation}
    \left\{
    \begin{alignedat}{9}
        &d_1 \, U_{zz} &&+ c \, U_z &&+ (r_1 &&- U &&- b_{12} \, V &&- b_{13} \, W &&)\,U &&= 0, \\
        &d_2 \, V_{zz} &&+ c \, V_z &&+ (r_2 &&- V &&- b_{21} \, U &&- b_{23} \, W &&)\,V &&= 0, \\
        &d_3 \, W_{zz} &&+ c \, W_z &&+ (r_3 &&- W &&- b_{31} \, U &&- b_{32} \, V &&)\,W &&= 0,
    \end{alignedat}
    \right.
    \label{eq:tw}
\end{equation}
with the asymptotic boundary conditions
\begin{equation}
    \begin{aligned}
         \lim_{z\to-\infty} (U, V, W)(z) &= (r_1, 0, 0), \\
         \lim_{z\to+\infty} (U, V, W)(z) &= (0, r_2, 0).
    \end{aligned}
    \label{eq:tw_bc}
\end{equation}
Clearly the unique solution of the two-species travelling wave problem given by \eqref{eq:tw2} and \eqref{eq:tw2_bc}
can be extended to a solution of problem \eqref{eq:tw} and \eqref{eq:tw_bc}
by setting $W \equiv 0$.
We call such a solution the \emph{trivial} travelling wave solution of \eqref{eq:cds:pde}.
Its velocity $c_{uv} > 0$
and its profile $(U,V)(z)$
are independent of the free parameter $r_3$.
Moreover, if $r_3$ is relatively small,
the trivial travelling wave is stable in the full three-species system \cite{ogawadrift}.
For the choice of parameters given in \eqref{eq:parameters},
we have $c_{uv} \approx 2.575$
and the wave profile is shown in Figure~\ref{fig:profiles} (top-left panel).

Does the problem \eqref{eq:tw} and \eqref{eq:tw_bc}
admit \emph{non-trivial} solutions
in which $W(z)$ is not zero everywhere?
Recently, it has been shown that non-trivial solutions never exist
if $r_3$ is sufficiently small \cite{nbarrier3}.
On the other hand, there are only a few analytical results
on the existence of non-trivial solutions \cite{chen}.
Numerically, the difficulty lies in finding the right parameter values for which non-trivial solutions exists.
Guided by our ecologically motivated assumptions and the values obtained in \intextcite{chen},
we have obtained the values shown in \eqref{eq:parameters}.
For such a choice of parameters, we have used the numerical continuation software AUTO \cite{auto}
to compute the global structure of the non-trivial solutions of \eqref{eq:tw} and \eqref{eq:tw_bc}
as the free parameter $r_3$ is varied.
In Figure~\ref{fig:bifurcation} we have plotted the resulting bifurcation diagram
for both trivial and non-trivial solutions of \eqref{eq:tw} and \eqref{eq:tw_bc}.

\ifchaos
\begin{figure*}
\else
\begin{figure}[p]
\fi
    \centering
    \includegraphics[width=0.85\textwidth]{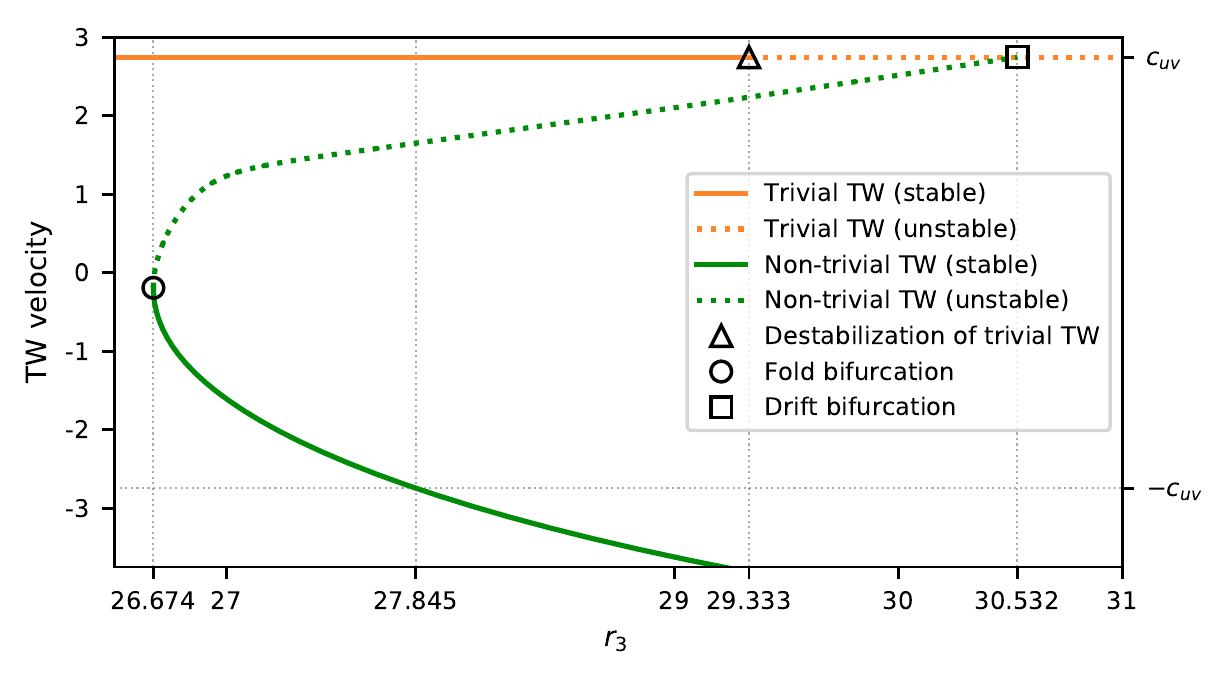}
    \caption{Bifurcation diagram for the trivial and non-trivial one-dimensional travelling wave solutions of \eqref{eq:cds:pde}.
             The travelling wave velocity is plotted as a function of the free parameter $r_3$.}
    \label{fig:bifurcation}
\ifchaos
\end{figure*}
\else
\end{figure}
\fi
\ifchaos
\begin{figure*}
\else
\begin{figure}[p]
\fi
    \centering
    \includegraphics[width=\textwidth]{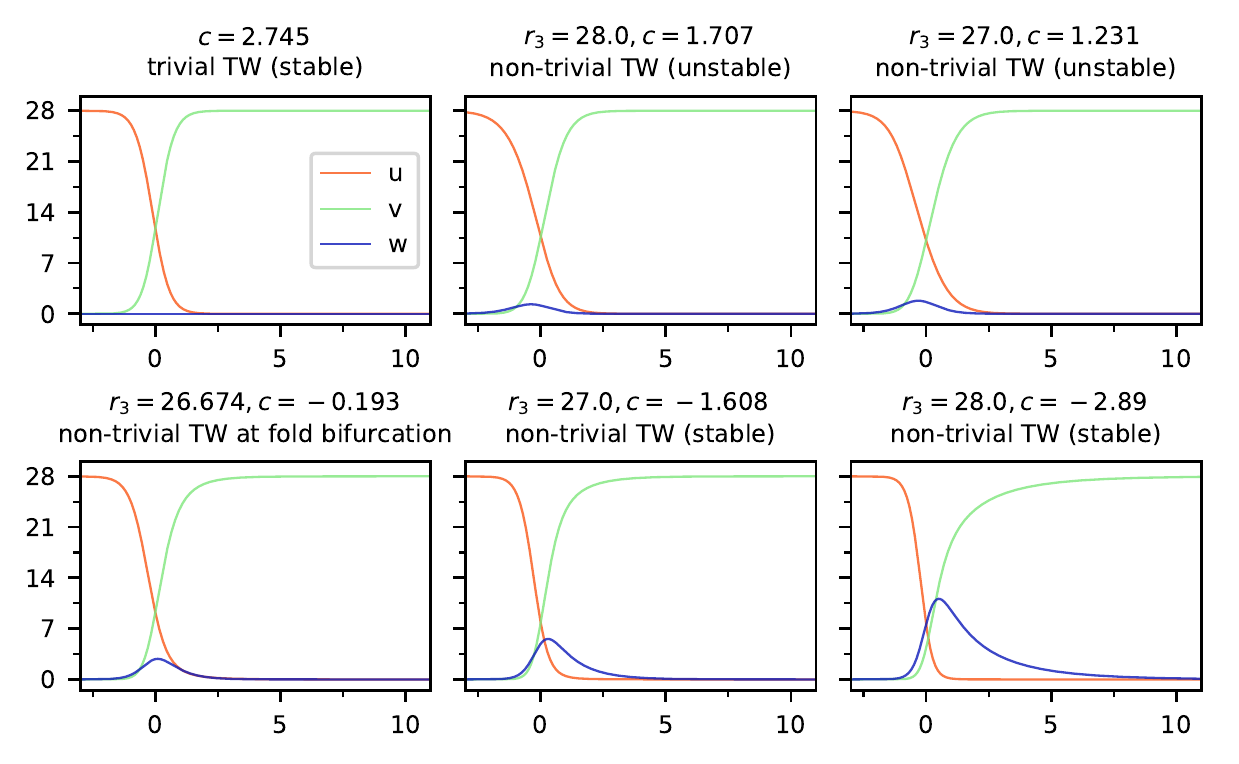}
    \caption{Profiles of the trivial and non-trivial one-dimensional travelling wave solutions of \eqref{eq:cds:pde} for different values of the free parameter $r_3$.}
    \label{fig:profiles}
\ifchaos
\end{figure*}
\else
\end{figure}
\fi

When $r_3$ is small there is no non-trivial travelling wave solution.
However, as $r_3$ increases, two non-trivial travelling waves appear
from a saddle-node bifurcation at $r_3 = r_F \approx 26.674$.
One of them (the lower branch) is stable with velocity $c_{uvw}$,
while the other (the upper branch) is unstable with velocity $\tilde{c}_{uvw}$.
The unstable branch eventually connects to the trivial travelling wave branch
at a \emph{drift bifurcation} point.

A drift bifurcation of a travelling wave occurs
when its zero eigenvalue is no longer simple.
A stable travelling wave has no eigenvalues with positive real part
and a simple zero eigenvalue.
This zero eigenvalue and the associated eigenfunction (Goldstone mode $G$,
given by the derivative of the profile with respect to the space variable)
are linked to the invariance of the traveling wave with respect to translations:
a small perturbation proportional to the Goldstone mode
will result in a shifted profile
which continues to move with the same shape and velocity.
However, as the parameters change,
an eigenvalue with zero imaginary part
may cross the imaginary axis.
Then, for a specific choice of the parameters
we have that zero is no longer a simple eigenvalue.
If it is also no longer regular,
we say that a drift bifurcation has occurred.
Then, in addition to the Goldstone mode,
there exists a generalized eigenfunction (propagator mode $P$)
which causes the traveling wave to destabilize.
In order to understand what happens to the destabilized wave,
let us consider how a small perturbation $\psi(t)$ to the wave profile evolves with time $t$.
The function $\psi$ satisfies the evolution equation given by
$\psi_t = \mathcal{L} \psi$,
where the linear operator $\mathcal{L}$
is obtained by
linearizing the travelling wave equation around the travelling wave profile.
The Goldstone and propagator modes satisfy $\mathcal{L} G = 0$ and $\mathcal{L} P = G$ by definition.
Supposing that the initial perturbation is proportional to the propagator mode,
i.e., $\psi(0) \propto P$,
we obtain that $\psi(t) \propto P + t G$.
Since an infinitesimal perturbation in the Goldstone mode is equivalent to shifting the wave profile,
the perturbed solution propagates with constant velocity in the reference frame of the original wave:
a new travelling wave, with a different profile and velocity, has been generated from the drift bifurcation.
For more details about drift bifurcations see, e.g., \cite{bode1997, gurevich2003}.
The existence of a drift bifurcation point
on the trivial travelling wave branch (with $r_3$ as the free parameter)
of a three-species competition-diffusion system
has been recently discussed in \cite{ogawadrift}.

Our case is slightly different from the simplest case presented above
(and also from \cite{ogawadrift}),
where the original travelling wave
loses stability at the drift bifurcation point
while the generated branch is stable.
Here instead, as can be seen in Figure~\ref{fig:bifurcation},
the trivial wave has already lost stability by the time it reaches the drift bifurcation point:
when $r_3$ crosses the threshold $29+1/3$, the species $v$ and $w$ become weakly competing
and the equilibrium $(0, r_2, 0)$ loses stability in the diffusion-free system.
The same discussion as above apply,
with the caveat that
the new travelling wave branches
generated at the drift bifurcation are all unstable.
Finally, we remark that
while in Figure~\ref{fig:bifurcation}
we have plotted the non-trivial travelling wave branch
only up to the drift bifurcation point,
this branch continues beyond it,
as can be expected from such a bifurcation.
However, the associated travelling wave solutions
are no longer admissible from a modelling point of view
since the density of $w$ around the front becomes negative.

In Figure~\ref{fig:profiles}
we show the evolution of the profile
of the non-trivial travelling wave
along its solution branch.
We start from the drift bifurcation point,
where a small peak of the invading species $w$
appears near the interface between the native species $u$ and $v$,
marking the origin of the non-trivial wave from the trivial one.
As we follow the branch, the peak gradually grows in size
and at the same time the velocity of the wave decreases,
until it becomes negative and the direction of wave propagation is inverted.
This is reasonable from a modelling point of view,
since by the cyclic competition property
the presence of the invader $w$ is harmful to $u$ and beneficial to $v$.
Thus, the speed of an interface between $u$ and $v$ is controlled by the density of $w$ around it.

We are also interested in seeing what happens if we perturb the trivial travelling wave
by increasing the density of $w$ near the interface.
For relatively small values of $r_3$,
the trivial wave is stable in the full three-species system,
so that small perturbations in $w$ decay back to zero.
If the perturbation in $w$ is above a certain threshold,
then the resulting profile will be more similar to a perturbed stable non-trivial wave.
As a consequence,
the interface will tend to that wave's shape and $w$ will be able to survive near the interface.
The unstable non-trivial wave acts like a \emph{separator} between these two different behaviours.
We remark that while single interfaces can only behave like either a trivial wave or a stable non-trivial wave,
when two or more interfaces interact they may behave differently and travel at other velocities,
as we will see in later sections.

We have seen that for a certain range of values of $r_3$
there exist \emph{two stable} front solutions
having the same asymptotic boundary conditions
but travelling in \emph{opposite} directions.
The same also occurs in FitzHugh-Nagumo-like systems
where the equivalent of a trivial wave (usually a standing wave)
destabilizes by a pitchfork (NIB) bifurcation
generating two symmetric branches of non-trivial stable travelling fronts
moving in opposite directions \cite{hagberg1994}.
We remark that such a bifurcation is also a drift bifurcation.
However, in that case the two stable waves are both non-trivial,
while in our case the trivial wave is stable and only one of the non-trivial waves is stable.
A situation much more similar to ours
has been recently discovered in a model of dryland vegetation \cite{zelnik2018}.
This model displays a one-species trivial wave describing the expansion of a herbaceous species
and a two-species non-trivial wave
characterized by the presence of a second species around the front.
While in this case the trivial wave is not independent of the free parameter
and the two waves propagate in the same direction,
the bifurcation diagram is remarkably similar to our case.

Finally, we stress that
when $d_1 = d_2 = d_3 = 1$,
as is the case in \eqref{eq:parameters},
both the trivial and the stable non-trivial one-dimensional travelling waves
are \emph{planarly stable} in two spatial dimensions,
i.e., their planar extensions are still stable.
Nevertheless the system displays very complex spatio-temporal patterns,
as shown in Figure~\ref{fig:r_3_mid}.
This is in opposition with otherwise similar systems,
where front instability is needed in order to observe complex behaviours
(e.g., in the case of fronts generated by NIB bifurcation,
spiral nucleation occurs in presence of front instability \cite{hagberg1997}).
While it is possible for the non-trivial wave
to become planarly unstable
if the diffusion rates are different enough,
in this paper we will not be concerned with such a case.

\section{Interaction of stable travelling waves in one dimension}
\label{sec:1d_tw_interaction}
As the starting point
for studying the complex spatio-temporal pattern shown in Figure~\ref{fig:r_3_mid},
we consider the interaction
of the two stable travelling wave solutions
with velocities $c_{uv}$ and $c_{uvw}$
in one spatial dimension.
Several outcomes are possible,
depending on the value of the free parameter $r_3$.

\begin{remark}
    \label{rem:swapped_bc}
    Strictly speaking, it is not possible
    to consider the interaction
    between the trivial and non-trivial waves introduced in Section~\ref{sec:1d_tw},
    since they both share the same asymptotic boundary conditions
    and such conditions differ for $z \to \pm \infty$:
    if we place them one next to the other the values at the contact point do not match.
    However, the travelling wave problem
    given by \eqref{eq:tw2} and \eqref{eq:tw2_bc}
    is invariant for the transformation $z \mapsto -z$, $c \mapsto -c$.
    Given a travelling wave solution, by reflecting its profile around the origin and changing its direction of propagation (but not the speed)
    we have a travelling wave solution which still satisfies \eqref{eq:tw2}
    with the swapped asymptotic boundary conditions
    \begin{equation}
        \begin{aligned}
            \lim_{z\to-\infty} (U, V, W)(z) &= (0, r_2, 0), \\
            \lim_{z\to+\infty} (U, V, W)(z) &= (r_1, 0, 0).
        \end{aligned}
        \label{eq:tw_swapped_bc}
    \end{equation}
    Then, the interaction of such a wave with another satisfying the boundary conditions \eqref{eq:tw2_bc}
    can be considered, since now the ends of the two waves match.
\end{remark}

First, we observe that if two waves of the same kind
travel in opposite directions
and collide with each other,
they are always annihilated.
The resulting uniform state is $(r_1, 0, 0)$ in the case of trivial waves \cite{morita}
or $(0, r_2, 0)$ in the case of non-trivial waves.

\ifchaos
\begin{figure*}
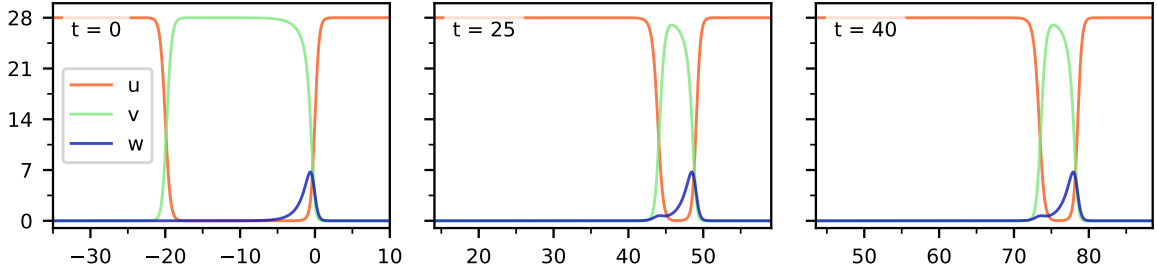

\else
\begin{figure}
\fi
    \centering
    \includegraphics[width=\textwidth]{{{tw_interaction__r3=27.2}}}
    \caption{Interaction of the trivial and non-trivial travelling waves for $r_3=27.2$, resulting in the waves merging into a single homoclinic travelling wave.}
    \label{fig:merging}
\ifchaos
\end{figure*}
\else
\end{figure}
\fi
\ifchaos
\begin{figure*}
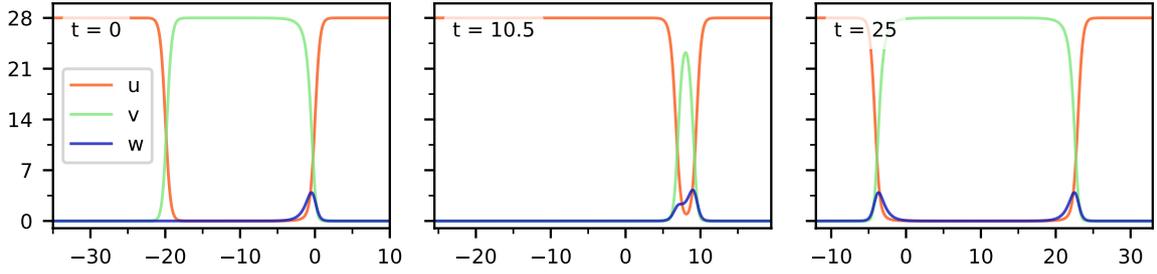

\else
\begin{figure}
\fi
    \centering
    \includegraphics[width=\textwidth]{{{tw_interaction__r3=26.75}}}
    \caption{Interaction of the trivial and non-trivial travelling waves for $r_3=26.75$,
             resulting in their reflection, with the trivial wave becoming a non-trivial wave moving in the opposite direction.}
    \label{fig:reflection}
\ifchaos
\end{figure*}
\else
\end{figure}
\fi

The interaction between trivial and non-trivial travelling waves yields more interesting results.
If $r_3$ is sufficiently large,
the non-trivial wave is the faster of the two.
This case and its consequences
for the existence of two-dimensional travelling wave solutions
have been discussed in \intextcite{contento}.
In this paper we are concerned only with values of $r_3$
for which the trivial wave is faster than the non-trivial wave.
Let us consider the situation in which both waves move towards the right,
with the the trivial wave approaching the non-trivial one from the left
so that they will collide eventually.
The outcome of their collision
depends on the relative speed $\Delta c = \lvert c_{uv} \rvert - \lvert c_{uvw} \rvert$ of the fronts,
which is itself a function of $r_3$.
In this and later sections,
whenever we consider two interacting interfaces
between regions where $u$ is dominant and regions where $v$ is dominant,
we call \emph{leading front} the interface most advanced in the direction of motion
and \emph{back front} the other.
In this case, initially the back front is the faster trivial wave,
while the leading front is the slower stable non-trivial wave.

If the leftmost trivial wave moves only slightly faster than the non-trivial wave,
then after collision the two waves merge and form a single travelling pulse,
i.e., a homoclinic travelling wave
with the same boundary value at both infinities.
An example of such an interaction is shown
in Figure~\ref{fig:merging}
for $r_3 = 27.2$,
which yields $c_{uvw} \approx -1.96$ and $\Delta c \approx 0.78$
(the corresponding animation can be found at \intextcite{movieTW1d}).
As the two fronts approach,
some of the invading species $w$ moves from the leading front to the back front.
Such an influx of $w$ is too small to pass the threshold beyond which the back front would become a non-trivial front,
but even a small presence of $w$ reduces the back front velocity, which becomes smaller than the original $c_{uv}$.
On the other hand, as a consequence of the decrease in $v$ due to the interfaces getting closer to each other,
$w$ grows in density also around the leading front, increasing its velocity from the original $c_{uvw}$, albeit only slightly.
Once the two fronts are sufficiently near each other,
these effects cancel their original speed difference
and they propagate like a single travelling pulse
at a velocity $c_h$ which is slightly larger than $c_{uvw}$.

If the relative speed is quite large instead,
then after collision the rightmost non-trivial wave
continues to propagate unaffected,
while the leftmost trivial wave changes nature,
becoming a second non-trivial wave moving in the opposite direction
as if it were reflected.
In this case the additional influx of $w$ from the leading front to the back front
is large enough for the back front to fully become a stable non-trivial front.
An example is presented in Figure~\ref{fig:reflection}
for $r_3 = 26.75$,
which yields $c_{uvw} \approx -0.9$ and $\Delta c \approx 1.85$
(the corresponding animation can be found at \intextcite{movieTW1d}).
We remark that most of the well-known cases of wave reflection
concern the interaction of travelling pulses (see, e.g., \intextcite{nagayama}),
which can reverse their direction of motion
by simple reflection of their profile
thanks to the fact that the boundary conditions at infinities are the same.
Since in our case the interaction occurs between travelling waves
with non-matching asymptotic boundary conditions,
the reversal of their direction of propagation can be achieved only
by a complete change of wave shape and velocity.
A similar reflection of fronts was observed in \cite[Figure~15(b)]{hagberg1994} in the context of NIB bifurcations
and showed two Bloch fronts of the same kind moving towards each other
and transforming into two rebounding Bloch fronts of the other type after collision.


\section{Homoclinic travelling waves in one dimension}
\label{sec:homoclinic}
In the previous section,
we observed that
the two different types of stable travelling waves of \eqref{eq:cds:pde}
may merge into a single homoclinic travelling wave with velocity $c_h$
if their speed difference is small.
The profile $(U,V,W)(z)$ of this homoclinic wave
solves the travelling wave equation \eqref{eq:tw} with $c = c_h$
and satisfies the asymptotic boundary conditions
\begin{equation*}
    \lim_{z\to\pm\infty} (U, V, W)(z) = (r_1, 0, 0).
\end{equation*}
Unfortunately, there is no analytical proof of the existence of this solution at the moment.
Therefore, we rely on AUTO to continue this solution numerically as $r_3$ varies,
starting from the profile obtained in the simulation of Figure~\ref{fig:merging}.
The resulting global structure
of the homoclinic travelling wave solution
is drawn in Figure~\ref{fig:bifurcation_homoclinic}.
As it was the case for the non-trivial wave,
its velocity $c_h$
depends on the free parameter $r_3$.
Spatial profiles of the homoclinic wave for some values of $r_3$ are plotted in Figure~\ref{fig:homoclinic_profiles}.

\ifchaos
\begin{figure*}
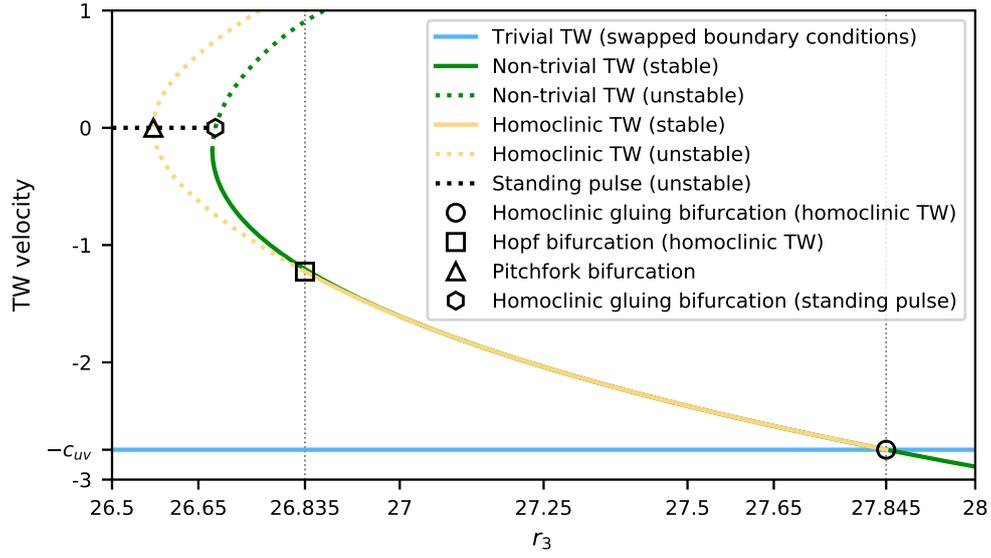

\else
\begin{figure}
\fi
    \centering
    \includegraphics[width=0.85\textwidth]{{{bifurcation_diagram_h}}}
    \caption{Bifurcation diagram for the one-dimensional travelling fronts and pulses of \eqref{eq:cds:pde}.
             The travelling wave velocity is plotted as a function of the free parameter $r_3$.}
    \label{fig:bifurcation_homoclinic}
\ifchaos
\end{figure*}
\else
\end{figure}
\fi
\ifchaos
\begin{figure*}
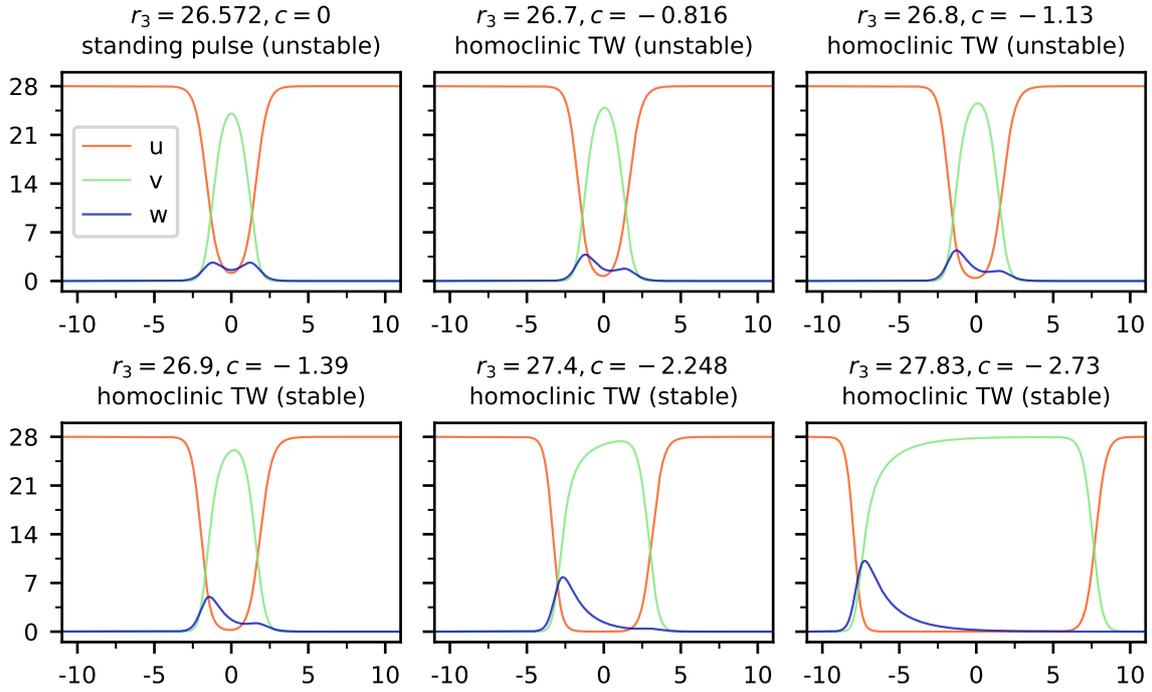

\else
\begin{figure}
\fi
    \centering
    \includegraphics[width=\textwidth]{{{h_tw_profiles}}}
    \caption{Profiles of the one-dimensional homoclinic travelling wave solutions of \eqref{eq:cds:pde},
             for different values of the free parameter $r_3$ and corresponding wave velocities $c$.
             Negative propagation velocities mean that the pulse is moving leftward.}
    \label{fig:homoclinic_profiles}
\ifchaos
\end{figure*}
\else
\end{figure}
\fi

On its right side,
the solution branch appears to start
from a homoclinic gluing bifurcation at $r_3 = r_G \approx 27.845$,
the value of $r_3$ for which
the trivial and non-trivial travelling waves
have the same speed,
i.e., $c_{uv} = - c_{uvw}$ and $\Delta c = 0$.
This bifurcation concerns the travelling waves
seen as heteroclinic/homoclinic orbits of a finite dimensional vector field
and not as solutions of a time-evolution problem for a reaction-diffusion system.
Moreover, strictly speaking, the waves under considerations
are not the trivial and the non-trivial waves:
the trivial wave must be substituted by the wave obtained by reflection around the origin,
as described in Remark~\ref{rem:swapped_bc}.
The velocity of such a wave is $-c_{uv}$
(which at the bifurcation point has the same sign \emph{and} modulus as $c_{uvw}$)
and its asymptotic boundary conditions are as in \eqref{eq:tw_swapped_bc}.

Then, at $r_3 = r_G$ we have two different heteroclinic orbits for the same set of parameters (\emph{including} the wave velocity),
both of them connecting the equilibria $(r_1, 0, 0)$ and $(0, r_2, 0)$
but oriented in different directions.
While there is no homoclinic orbit connecting $(r_1, 0, 0)$ with itself at $r_3 = r_G$,
under some hypotheses a branch of such solutions will be born on one side of the bifurcation point
(for mathematical details concerning this and other related bifurcations,
such as the homoclinic doubling bifurcation, we refer to \cite{kokubo}).
This homoclinic orbit corresponds to the travelling pulse shown in Figure~\ref{fig:homoclinic_profiles}.
Moreover, since near the bifurcation point
the homoclinic orbit must closely follow the two heteroclinic orbits,
for values of $r_3$ very close to $r_G$
the homoclinic wave profile looks like a trivial wave glued together with a non-trivial wave,
as shown in Figure~\ref{fig:homoclinic_profiles} (bottom-right panel).
Such a gluing of waves is already well known in the literature.
In particular,
in \cite{kokubopitch, hagberg1993, ikedaglue}
the gluing of the two non-trivial fronts originating from NIB bifurcation in FitzHugh-Nagumo-like systems is studied,
in \cite{sandstedeglue} the gluing of two unstable fronts is shown to produce a stable pulse,
and in \cite{goh} the more complicated gluing of a travelling front with a pattern-forming front is considered.
Finally, we remark that,
in accordance to the properties of homoclinic gluing bifurcations,
we expect a second homoclinic orbit
to be generated at $r_3 = r_G$.
This second orbit should connect $(0, r_2, 0)$ with itself,
but we believe the associated travelling pulse to be unstable,
since we were not able to observe it numerically.
Further study is needed in order to uncover its nature and global structure.

As we have observed in the previous section about travelling wave interaction
and can also be seen clearly from Figure~\ref{fig:homoclinic_profiles} (bottom-right panel),
the pulse is composed
of two distinct interfaces between regions
where either $u > v$ or $u < v$,
the leading and the back fronts
(in Figure~\ref{fig:homoclinic_profiles} the leading front is the leftmost one
since the velocity of propagation $c_h$ is negative).
Near the homoclinic gluing bifurcation,
the leading front resembles the slower non-trivial wave,
while the back front resembles the faster trivial wave (with swapped boundary conditions; see Remark~\ref{rem:swapped_bc}).
However, due to immigration from the leading front,
a small amount of the invading species $w$ is present also around the back front,
reducing its velocity and preventing it to collide with the leading front.
As $r_3$ decreases the speed difference between the trivial and the non-trivial waves increases,
so that the back front must get closer to the leading front
to allow the influx of $w$ to be strong enough to equilibrate the velocities of the two fronts.
This means that, as we get further away from the homoclinic gluing bifurcation,
the width of the pulse becomes smaller
and the peak of $w$ around the back front becomes larger,
as can be seen from the wave profiles in Figure~\ref{fig:homoclinic_profiles}.
Moreover, as the pulse width becomes smaller,
the density of the species $v$ around the leading front decreases.
Since $v$ is stronger than $w$ in the competition as per hypothesis \eqref{eq:A3_vw_mono},
its decay allows the invader $w$ near the leading front
to grow more than it would be possible in a normal non-trivial front.
Since to higher densities of $w$ near the interface it corresponds a faster propagation speed,
the velocity of the leading front, and thus of the homoclinic travelling wave,
is always greater in modulus than the velocity of the stable non-trivial wave,
i.e., $c_h < c_{uvw}$.
Since the pulse width shrinks as $r_3$ decreases,
the difference between the velocities of the homoclinic and stable non-trivial waves
increases as $r_3$ decreases,
as can be seen in the bifurcation diagram of Figure~\ref{fig:bifurcation_homoclinic}.

As for the stability of the homoclinic travelling wave,
for values of $r_3$ sufficiently near to $r_G$
it can be verified numerically that the wave is stable.
These are the values of $r_3$ for which the collision of trivial and non-trivial waves
results in their merging into a single pulse, as previously shown in Figure~\ref{fig:merging}.
However, at a certain point $r_3 = r_H \approx 26.83586$
the homoclinic wave goes through a Hopf bifurcation
and loses stability.
The behaviour of travelling wave interaction after the Hopf bifurcation
will be studied later in Section~\ref{sec:reflection_mechanism}.

Once $r_3$ reaches the value $r_3 = r_P \approx 26.572$,
the velocity of the travelling pulse becomes zero
and the corresponding wave profile is symmetric,
as shown in Figure~\ref{fig:homoclinic_profiles} (top-left panel).
This point is a pitchfork bifurcation,
where the asymmetric travelling pulse
branches off a symmetric standing pulse solution
(i.e., a travelling pulse with zero velocity).
Since for a pulse the boundary conditions at infinities are the same,
on the other side of the standing wave branch
there is a second asymmetric travelling pulse branch,
which is exactly the reflection of the first one around the axis $c = 0$.

As for the standing pulse solution,
we believe that
it is generated at $r_3 = r_{G'} \approx 26.679$
from a degenerate homoclinic gluing bifurcation
associated to two zero-velocity unstable non-trivial travelling waves
oriented in opposite directions.
Its initial profile can be seen in Figure~\ref{fig:standing_profiles} (rightmost panel).
Note that the point at which the non-trivial wave has zero velocity
\emph{is not} the fold bifurcation of its branch.
As $r_3$ decreases, the pulse shrinks,
as shown in Figure~\ref{fig:standing_profiles} (third and second panels),
until its profile is no longer non-negative
and the wave becomes non-admissible from a modelling point of view,
as shown in Figure~\ref{fig:standing_profiles} (leftmost panel).
All along its branch, this standing wave solution is unstable.
Probably for this reason, it does not appear to play an important role
in the patterns displayed by this system
and we will not be concerned with it any further in this paper.

\ifchaos
\begin{figure*}
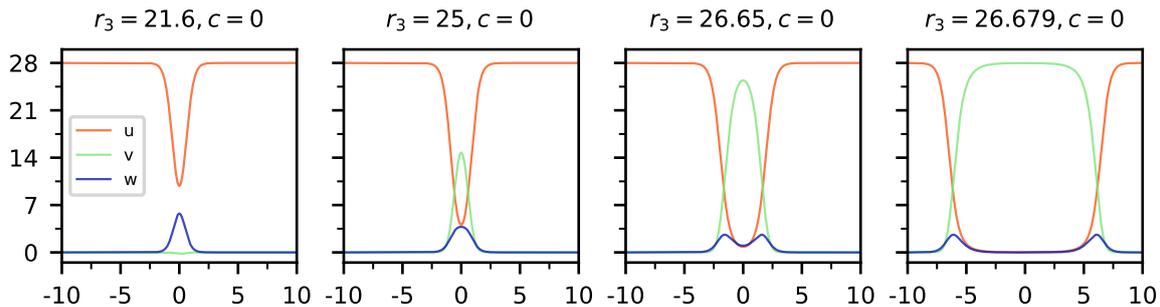

\else
\begin{figure}
\fi
    \centering
    \includegraphics[width=\textwidth]{{{standing_pulse_profiles}}}
    \caption{Profiles of the one-dimensional standing pulse solutions of \eqref{eq:cds:pde},
             for different values of the free parameter $r_3$.
             Note that in the leftmost panel the species $v$ assumes negative density values
             and the pulse is thus inadmissible from a modelling point of view.}
    \label{fig:standing_profiles}
\ifchaos
\end{figure*}
\else
\end{figure}
\fi

\section{Breathing travelling waves}
\label{sec:breathing}
In the previous section we showed how the homoclinic travelling wave solution
is destabilized through a Hopf bifurcation,
as shown in the bifurcation diagram of Figure~\ref{fig:bifurcation_homoclinic}.
When a dynamical system goes through a Hopf bifurcation
a periodic orbit is generated
and this section will be devoted to studying such a periodic solution and its bifurcation structure.

Let us consider
the time-evolution problem associated
to the travelling wave equation \eqref{eq:tw}:
\begin{equation}
    \left\{
    \begin{alignedat}{7}
        U_\tau &= d_1 \, U_{zz} &&+ c \, U_z &&+ (r_1 &&- U &&- b_{12} \, V &&- b_{13} \, W &&)\, U, \\
        V_\tau &= d_2 \, V_{zz} &&+ c \, V_z &&+ (r_2 &&- V &&- b_{21} \, U &&- b_{23} \, W &&)\, V, \\
        W_\tau &= d_3 \, W_{zz} &&+ c \, W_z &&+ (r_3 &&- W &&- b_{31} \, U &&- b_{32} \, V &&)\, W,
    \end{alignedat}
    \right.
    \label{eq:cds_moving_frame}
\end{equation}
where the profile $(U,V,W)(z,\tau)$ now also depends on a time variable $\tau$.
We remark that \eqref{eq:cds_moving_frame}
is equivalent to considering
the competition-diffusion system \eqref{eq:cds:pde}
on the real line
and in a reference frame moving with velocity $c$.
Then, the profile of the homoclinic travelling wave of \eqref{eq:cds:pde}
is a stationary solution of \eqref{eq:cds_moving_frame}
when $c = c_h$, with $c_h$ being the velocity of the homoclinic wave.
As we have seen in the previous section,
at $r_3 = r_H$ this solution
goes through a Hopf bifurcation and loses stability.
Then, we can expect that a time-periodic solution of \eqref{eq:cds_moving_frame}
is generated on one side with respect to $r_H$.
We will denote by $T$ the period of such a solution
and by $c_b$ the velocity of the reference frame;
both quantities depend on the free parameter $r_3$.
The initial values of $c_b$ and $T$ at $r_3=r_H$
are expected to be equal to $c_h$ and $2\pi / \omega \approx 5.377$ respectively,
where $\omega \approx 1.169$ is the modulus of the imaginary part of the couple of eigenvalues crossing the imaginary axis at the Hopf bifurcation.

In the fixed frame of reference,
this solution appears like a travelling wave
whose profile oscillates periodically.
It can be written as
$(u,v,w)(x,t) = (U,V,W)(x - c_b \, t, t)$,
where $c_b$ is the wave velocity
and $(U,V,W)(z,\tau)$ is the travelling wave profile
which satisfies \eqref{eq:cds_moving_frame} with $c=c_b$
and is $T$-periodic in the time variable $\tau$.
In particular, in this case it is the pulse width that oscillates periodically,
which earned the name of \emph{breathing travelling waves} (or \emph{travelling breathers})
for solutions of this type (see, e.g., \intextcite{mimuraBreathing}).
If the wave velocity $c_b$ is zero, then we have standing breathers
such as those observed in the case of a FitzHugh-Nagumo-like system with NIB bifurcation \cite[Figures~14 and 15(a)]{hagberg1994}.
We remark that $c_b$ is the mean velocity of the wave
over time-intervals of length $T$.
The instantaneous front velocity may differ from $c_b$
and is in general a $T$-periodic function of time.

\ifchaos
\begin{figure*}
\else
\begin{figure}[p]
\fi
    \centering
    \includegraphics[width=\textwidth]{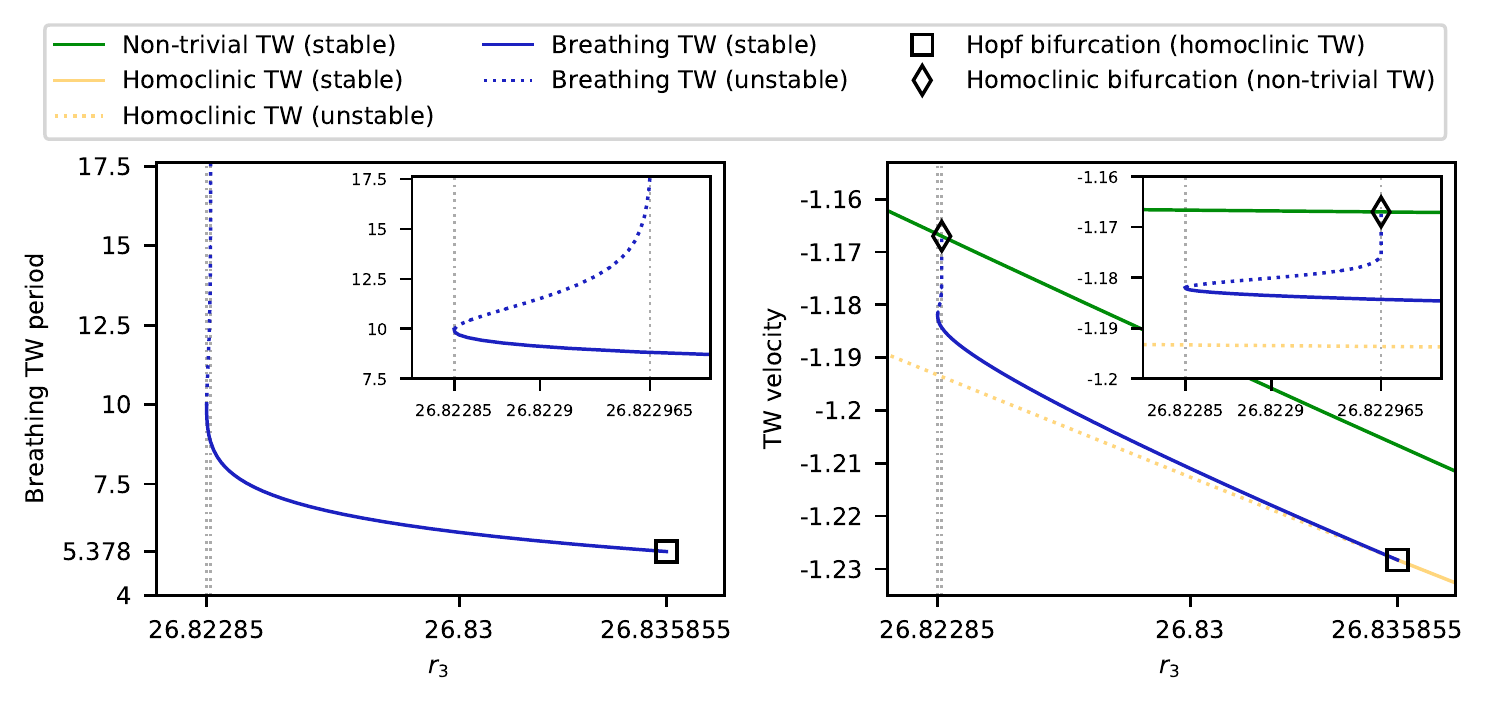}
    \caption{Global structure of the breathing travelling wave solution.
             Its period $T$ (left panel) and its mean velocity $c_b$ (right panel)
             have been plotted
             as functions of the free parameter $r_3$.
             Details of the unstable branch are also present.}
    \label{fig:breathing_period_plot}
\ifchaos
\end{figure*}
\else
\end{figure}
\fi
\ifchaos
\begin{figure*}
\else
\begin{figure}[p]
\fi
    \centering
    \includegraphics[width=\textwidth]{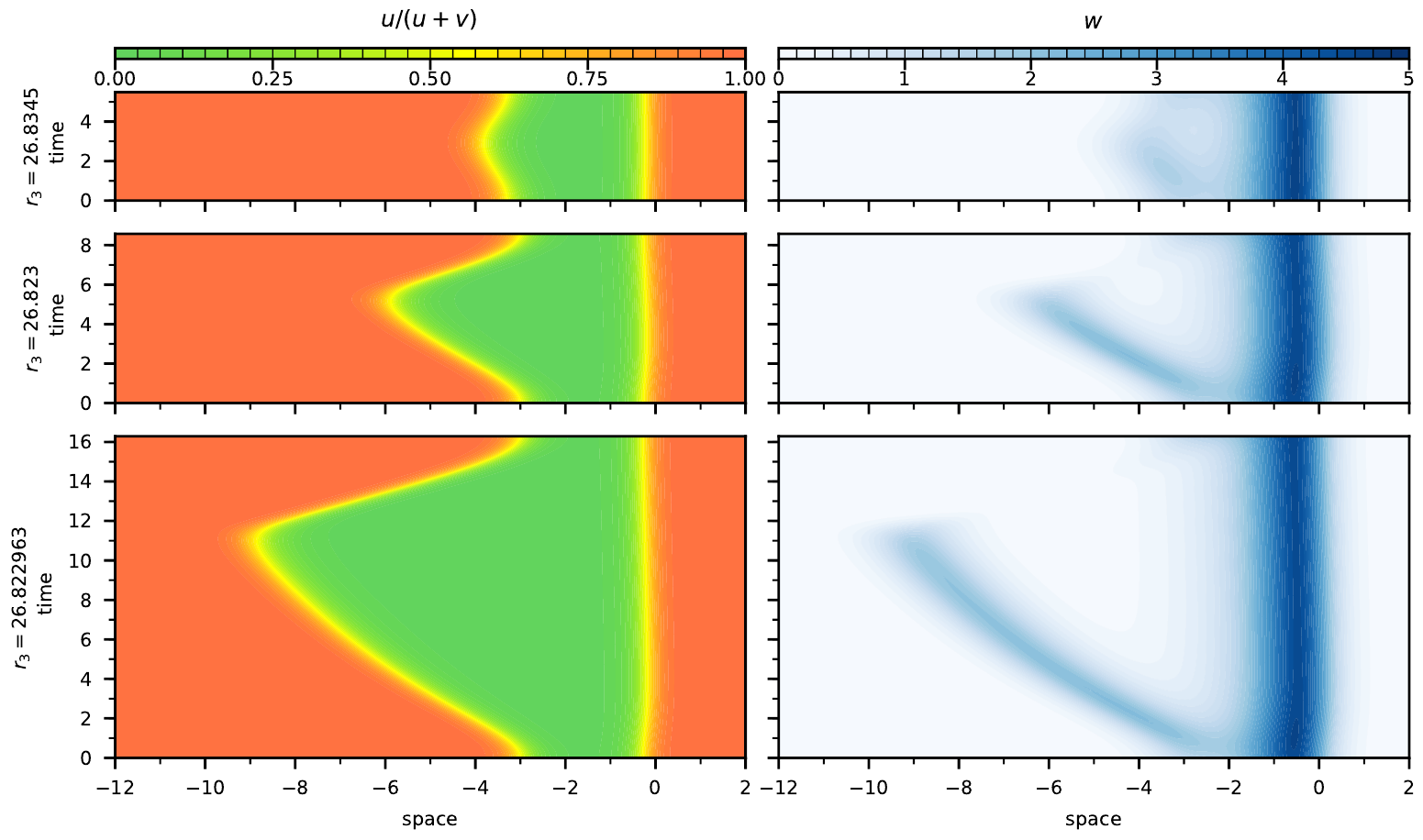}
    \caption{Space-time profiles of the breathing travelling wave over one time period $T$ for three different values of $r_3$.
             Instead of plotting $u$ and $v$ separately,
             we have used $u/(u+v)$ as a measure of the relative strength of $u$ and $v$.
             The spatial frame of reference moves at the velocity $c_b$, the mean velocity of the breathing wave.}
    \label{fig:breathing_spacetime}
\ifchaos
\end{figure*}
\else
\end{figure}
\fi
\ifchaos
\begin{figure*}
\else
\begin{figure}[p]
\fi
    \centering
    \includegraphics[width=\textwidth]{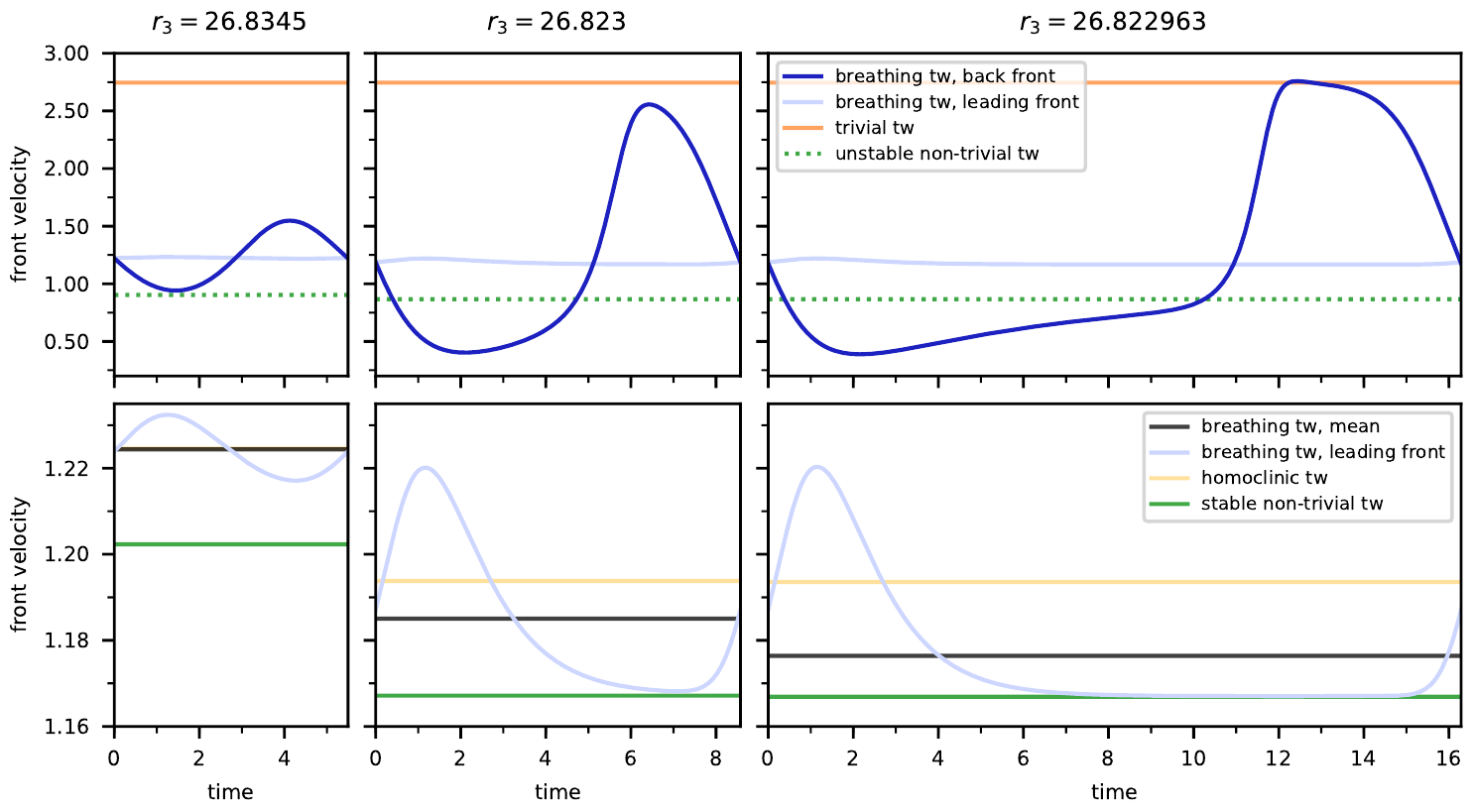}
    \caption{Temporal profiles of the instantaneous velocities of the back and leading fronts
             of the breathing travelling wave
             for three different values of $r_3$.
             The unsigned speeds of the other types of travelling waves are also plotted for comparison.}
    \label{fig:breathing_velocities}
\ifchaos
\end{figure*}
\else
\end{figure}
\fi
\ifchaos
\begin{figure*}
\else
\begin{figure}[p]
\fi
    \centering
    \includegraphics[width=\textwidth]{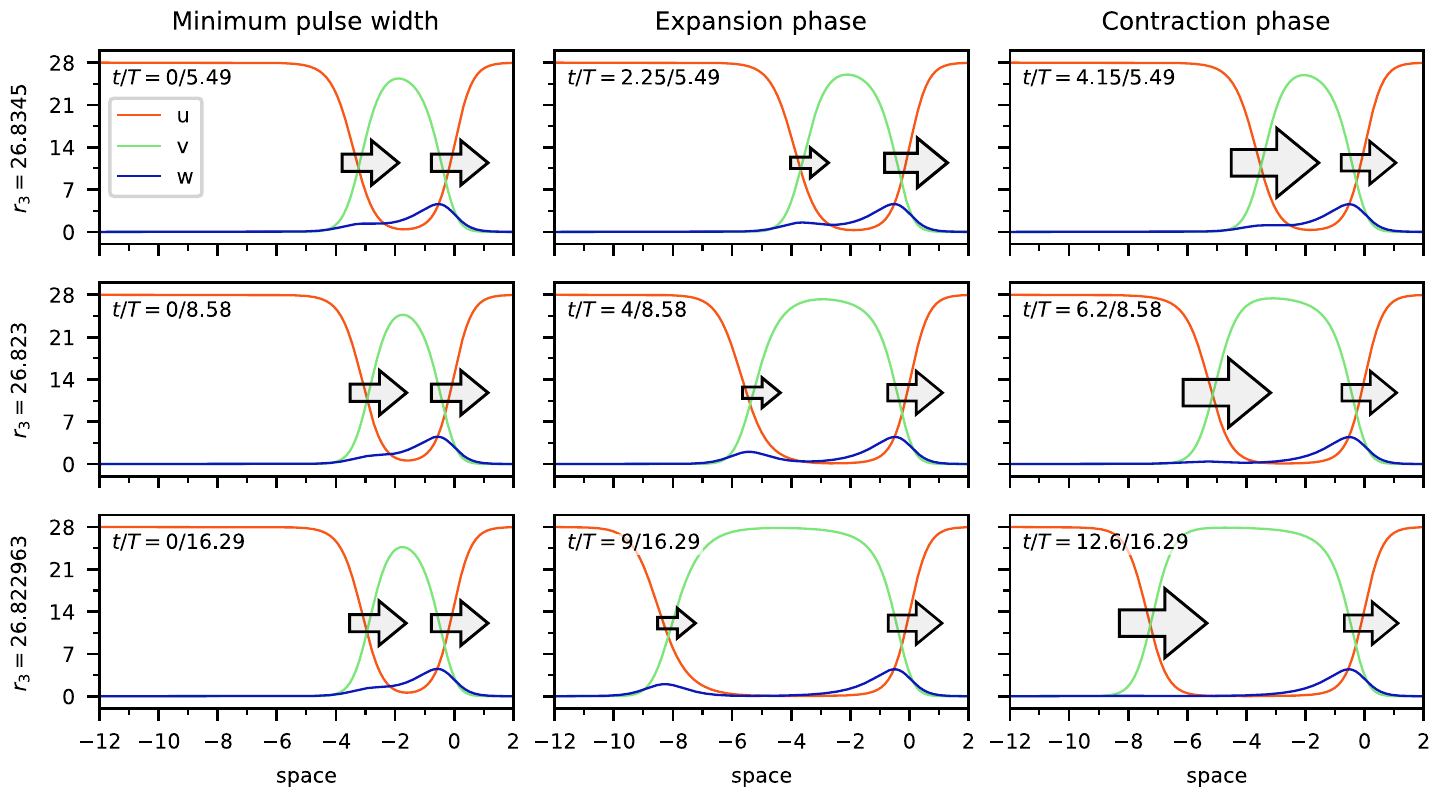}
    \caption{Spatial profiles of the breathing travelling wave
             during the expansion and contraction phases
             for three different values of $r_3$.
             The spatial frame of reference moves at the velocity $c_b$, the mean velocity of the breathing wave.
             The arrows show the direction of propagation and the speed of the two fronts.}
    \label{fig:breathing_profiles}
\ifchaos
\end{figure*}
\else
\end{figure}
\fi

In Figure~\ref{fig:breathing_period_plot}
we plot the period $T$ and mean velocity $c_b$ of the breathing wave as functions of $r_3$,
as obtained by numerical continuation.
The period $T$ seems to go to infinity at one end of the branch,
making the numerical solution increasingly difficult to obtain.
Moreover, we observe that it holds $\lvert c_h \rvert > \lvert c_b \rvert > \lvert c_{uvw} \rvert$,
with $c_b$ starting equal to $c_h$ near the Hopf bifurcation point
and apparently tending to $c_{uvw}$ as the period $T$ goes to infinity.
As some concrete examples of the breathing travelling wave,
we have chosen three values of $r_3$
and reported the features of the corresponding breathing solutions
in Figures~\ref{fig:breathing_spacetime}, \ref{fig:breathing_velocities} and \ref{fig:breathing_profiles}.
Animations containing the same features for a larger number of values of $r_3$
can be found online at \intextcite{movieBreathing}.
First, we focus on the general properties of the breathing wave, shared by all values of $r_3$.
Later we will describe how the breathing wave gradually changes as the free parameter $r_3$ is decreased.

In Figure~\ref{fig:breathing_spacetime}
we have plotted the breathing wave profile $(U,V,W)(z,\tau)$
over one time period $T$.
Since $u$ and $v$ are essentially complementary,
we have used the relative strength indicator $u/(u+v)$
instead of showing both $u$ and $v$ separately.
Moreover, since the spatial frame of reference used is moving at the velocity $c_b$,
globally the wave appears to be still.
From Figure~\ref{fig:breathing_spacetime}
we can see that also the breathing wave is composed by a back and a leading front
(here the leftmost and rightmost fronts respectively),
similarly to the homoclinic wave from which it originates.
The mean velocity of these two fronts over time intervals of length $T$ is equal to $c_b$,
but their instantaneous velocities
are not constant and are plotted in Figure~\ref{fig:breathing_velocities}.
As a means of comparison,
the \emph{unsigned} speeds of the other travelling waves are also plotted.
For the fronts of the breathing wave the signed velocity is instead used,
showing that they are always moving in the same direction.
In Figure~\ref{fig:breathing_spacetime}
the back front appears to move backwards to the left for some time,
but this motion is relative to the reference frame moving with the mean breathing wave velocity $c_b$.
In the fixed reference frame both the leading and the back fronts are always moving forward to the right.

The breathing wave cycle can be divided into two phases.
In the first one, the back front is slower than the leading front,
thus causing the distance between the fronts, i.e., the pulse width, to increase.
The second phase sees the back front becoming faster than the leading front
and the pulse width shrinking as a consequence.
This is possible since the amount of the species $w$ around the back front
appears to be greatly reduced compared to the first phase
(especially as we get further away from the Hopf bifurcation),
thus increasing the back front velocity.
Also note that
both the maximum pulse width and the period $T$
increase as we move along the solution branch.

Compared to the back front,
the instantaneous velocity of the leading front oscillates around $c_b$ weakly.
In order to appreciate its oscillation,
we have provided enlargements in the second row of Figure~\ref{fig:breathing_velocities}.
The maximum velocity is achieved some time after the minimum pulse width has been attained,
because, due to the small pulse width, the density of the species $v$ around the leading front is reduced.
Since $v$ is stronger than $w$ as per hypothesis \eqref{eq:A3_vw_mono},
its reduction favours the invading species $w$
and brings about a small increase in its density around the leading front,
which then becomes slightly faster.

Finally, in Figure~\ref{fig:breathing_profiles}
we show the spatial profile of the breathing wave
at three different time instants.
The first column shows the time at which the pulse width is the shortest.
Note how at this time the velocities of the leading and back fronts are the same
and how the wave profile is quite similar to that of the homoclinic wave.
In the second column, a snapshot of the expansion phase of the pulse width is shown,
where the velocity of back front becomes smaller than the velocity of the leading front.
In the last column, the contraction phase is depicted,
in which the back front becomes faster than the leading front.

Now, we follow the evolution of the breathing travelling wave along its solution branch,
as plotted in Figure~\ref{fig:breathing_period_plot}.
We start from the vicinity of the Hopf bifurcation point,
for example taking $r_3 = 26.8345$,
for which $T \approx 5.49$ and $c_b \approx 1.224$,
as shown in the first row of
Figures~\ref{fig:breathing_spacetime} and \ref{fig:breathing_profiles}
and in the first column of Figure~\ref{fig:breathing_velocities}.
In this case the breathing wave does not deviate too much from the homoclinic wave,
either in profile or velocity.
The back front velocity does not oscillate very far from the mean velocity $c_b$
and the reduction of the density of $w$ around the back front during the contraction phase
is not very significant.
In particular, at all times the area where $w$ is non-negligible
straddles both fronts, connecting them as in the homoclinic wave.

As $r_3$ decreases,
the maximum pulse width increases
and so does the time-period $T$.
As an example, we consider $r_3 = 26.823$,
for which $T \approx 8.58$ and $c_b \approx 1.185$,
as shown in the second row of
Figures~\ref{fig:breathing_spacetime} and \ref{fig:breathing_profiles}
and in the second column of Figure~\ref{fig:breathing_velocities}.
Now we will describe in detail the behaviour of this solution for one time period.

At the instant of shortest pulse width,
the breathing wave profile is still very similar to the homoclinic wave
and the same can be said for the velocity of both back and leading fronts.
Then, the pulse width increases
and compared to the previous case ($r_3=26.8345$)
the two interfaces travel further away from each other
and the two peaks of $w$ around them become more clearly separated.

In the expansion phase, the leading front looks quite similar
to the stable non-trivial travelling wave with velocity $c_{uvw}$.
Just after the minimum pulse width is reached,
its speed is slightly larger than $\lvert c_{uvw} \rvert$,
and even larger than the homoclinic wave speed $\lvert c_h \rvert$,
due to the reduced density of $v$ caused by the small pulse width.
As the distance between the fronts increases
this effect becomes weaker,
so that velocity and profile of the leading front
approach those of the stable non-trivial wave.
On the other hand, in the expansion phase
the back front resembles the unstable non-trivial travelling wave.
Similarly to the leading front,
at the beginning of the expansion phase
the peak value of $w$ near the back interface is slightly higher
than that of the unstable non-trivial wave,
but as time progresses it decays slowly
and the velocity of the back front tends to that of the unstable non-trivial wave.

Due to the unstable nature of such a wave,
at a certain point
the species $w$ around the back front dies out quite abruptly.
As a consequence, the back front accelerates, becoming faster than the leading front
and marking the beginning of the contraction phase of the pulse width.
In this phase, the back front becomes similar in appearance and velocity to the trivial travelling wave,
while the leading front is still resembling the stable non-trivial wave.
When the two fronts are sufficiently near for $w$ to colonize again the back front,
the back front slows down and matches in speed the leading one,
bringing us back to the starting point.
Thus, the behaviour of the breathing wave
may also be described as some kind of imperfect or incomplete reflection:
a faster trivial wave collides with a slower stable non-trivial wave,
then is ``reflected'' without changing direction of motion
as an even slower unstable non-trivial wave,
eventually decays back to a trivial wave
and the cycle starts again from the beginning.

As $r_3$ continues to decrease,
the back front travels further backward before destabilizing,
leading to an increase in period and maximum pulse width.
At $r_3 = r_{BF} \approx 26.82285$,
the breathing travelling wave undergoes a fold bifurcation and disappears.
However, using numerical continuation we can follow the unstable branch,
for which the period $T$ increases as $r_3$ increases,
as can be seen in the detail panels of Figure~\ref{fig:breathing_period_plot}.
Solutions on the unstable branch have the same qualitative behaviour as those on the stable branch,
but the maximum front separation and increase in period become even more dramatic.
As an example, we consider the unstable breathing wave for $r_3 = 26.822963$,
with $T \approx 16.29$ and $c_b \approx 1.176$,
as shown in the third row of
Figures~\ref{fig:breathing_spacetime} and \ref{fig:breathing_profiles}
and in the third column of Figure~\ref{fig:breathing_velocities}.
Note how the decay in the density of $w$ around the back front
during the expansion phase is much slower,
allowing the fronts to travel further away from each other
until they look like two separate waves.
The fraction of the period
in which the back front resembles
either the unstable non-trivial wave or the trivial wave
becomes larger.
Similarly, the leading front behaves like the stable non-trivial wave for a longer time
and consequently the mean breathing wave velocity will get away from the homoclinic wave velocity
and approach the velocity of the stable non-trivial wave,
as can be seen in the right panel of Figure~\ref{fig:breathing_period_plot}.

While numerical computation becomes increasingly difficult as the time-period $T$ increases,
it appears that $T$ tends to infinity as $r_3$ tends to a certain value $r_{HC} \approx 26.822965$.
Since the maximum pulse width increases together with the period,
we expect the maximum distance between the fronts to also tend to infinity,
while the breathing wave velocity $c_b$ should tend to $c_{uvw}$
since the leading front behaves like a stable non-trivial wave for most of the time.
Since the period goes to infinity,
the time-periodic breathing travelling wave solution
tends to an entire solution of \eqref{eq:cds_moving_frame},
i.e., a solution defined for all times $\tau\in\mathbb{R}$.
Moreover,
since the two fronts get arbitrarily far away from each other,
it appears that this entire solution describes
the interaction of
a faster trivial wave
with a slower stable non-trivial wave.
After interaction,
the back front slows down without changing direction,
its shape and velocity tending to those of an unstable non-trivial wave
as the time goes to infinity.
On the other hand,
the leading stable non-trivial wave is essentially unaffected by the collision.
In particular,
since the moving frame is centred on the leading front,
in the entire solution the spatial profile will tend pointwise
to the profile of the stable non-trivial wave
as the time goes to $\pm \infty$.
In other words, for $r_3 = r_{HC}$
it seems that we have a homoclinic bifurcation in the system \eqref{eq:cds_moving_frame},
for the equilibrium given by the stable non-trivial travelling wave,
the homoclinic orbit at $r_3 = r_{HC}$ given by the entire solution described above
and the periodic orbit for $r_3 < r_{HC}$ given by the breathing travelling wave.

As we will see in the next section,
in numerical simulations
the interaction of a trivial wave with a stable non-trivial wave
either results in two stable non-trivial waves for $r_3 < r_{HC}$
or in a single pulse for $r_3 > r_{HC}$.
This suggests that
the entire solution associated to the homoclinic bifurcation
must be unstable,
since we never observe the back front becoming an unstable non-trivial wave.
Confirming the existence of such a solution at $r_3 = r_{HC}$
and studying its evolution as $r_3$ is varied
are still open problems,
both analytically and numerically.
We believe that
reducing the wave interaction problem
to a system of ordinary differential equations
by a singular limit procedure
may allow the problem to become more tractable.

Finally, for a description of a similar bifurcation structure
for a different type of time-periodic solutions occurring in another reaction-diffusion system,
along with details about the numerical methods needed for the numerical continuation of such solutions,
we refer the interested reader to \intextcite{yadomeBifurcation}.

\section{The mechanism behind travelling wave reflection}
\label{sec:reflection_mechanism}
Going back to the problem of one-dimensional travelling wave interaction
presented in Section~\ref{sec:1d_tw_interaction},
we discover that the homoclinic and breathing travelling wave solutions
we have studied in Sections~\ref{sec:homoclinic} and \ref{sec:breathing} respectively
are deeply linked with the transition from the merging-after-collision regime of Figure~\ref{fig:merging}
to the reflection regime of Figure~\ref{fig:reflection}.
This is not completely unexpected,
since for example it is known that unstable twin-peaked solutions called scattors
have a fundamental role in the interaction of travelling pulses
in the Gray-Scott model \cite{nishiuraInteraction}.

First, we slightly perturb the homoclinic travelling wave
and see how the resulting wave evolves with time.
Clearly, if $r_3$ is larger than the Hopf bifurcation point $r_H$,
then the homoclinic wave is stable:
the perturbation simply decays to zero and the wave reverts to its original shape.
If $r_{BF} < r_3 < r_H$ the homoclinic wave is unstable,
but a stable breathing wave does exist.
Then, in this parameter range,
the distance between the back and leading fronts of the perturbed pulse starts oscillating
and the solution tends to the breathing wave.
If $r_F < r_3 < r_{BF}$ no breathing wave exists
and the back front, after oscillating for a while, reverses its direction of propagation.
The end result consists in two stable non-trivial waves moving in opposite directions,
as can be seen in Figure~\ref{fig:homoclinic_destabilization:splitting} for the example value $r_3 = 26.75$
(the corresponding animation can be found at \intextcite{moviePerturbation}).
If $r_3 < r_F$ then the non-trivial fronts no longer exist
and the homoclinic wave decays to the homogeneous state $(r_1,0,0)$.

\ifchaos
\begin{figure*}
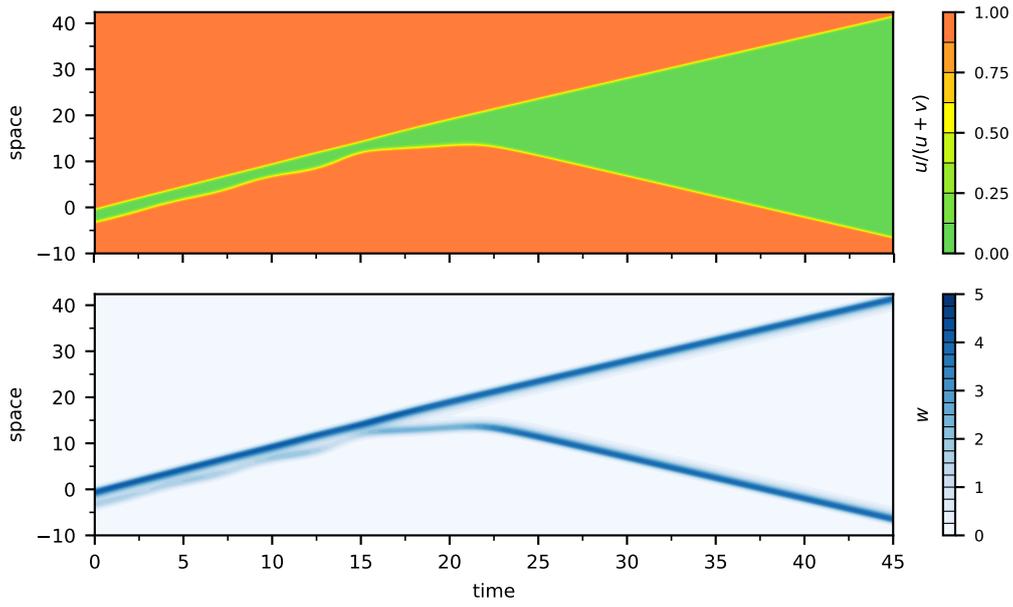

\else
\begin{figure}[p]
\fi
    \centering
    \includegraphics[width=0.87\textwidth]{{{pulse_destab__r3=26.75000__spacetime}}}
    \caption{Perturbing an unstable homoclinic wave for $r_3=26.75$ results in two stable non-trivial fronts moving in opposite directions.}
    \label{fig:homoclinic_destabilization:splitting}
\ifchaos
\end{figure*}
\else
\end{figure}
\fi
\ifchaos
\begin{figure*}
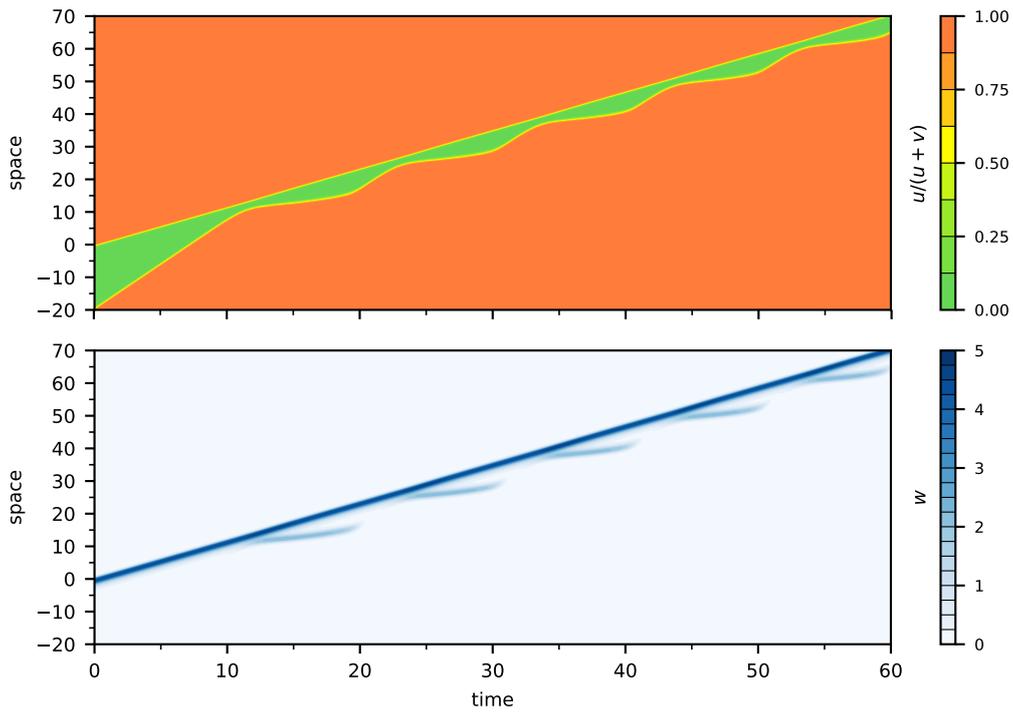

\else
\begin{figure}[p]
\fi
    \centering
    \includegraphics[width=0.87\textwidth]{{{tw_interaction__r3=26.823__spacetime}}}
    \caption{Interaction of the trivial and non-trivial travelling waves for $r_3=26.823$, resulting in the waves merging into the stable breathing travelling wave.}
    \label{fig:merging_to_breathing}
\ifchaos
\end{figure*}
\else
\end{figure}
\fi
\ifchaos
\begin{figure*}
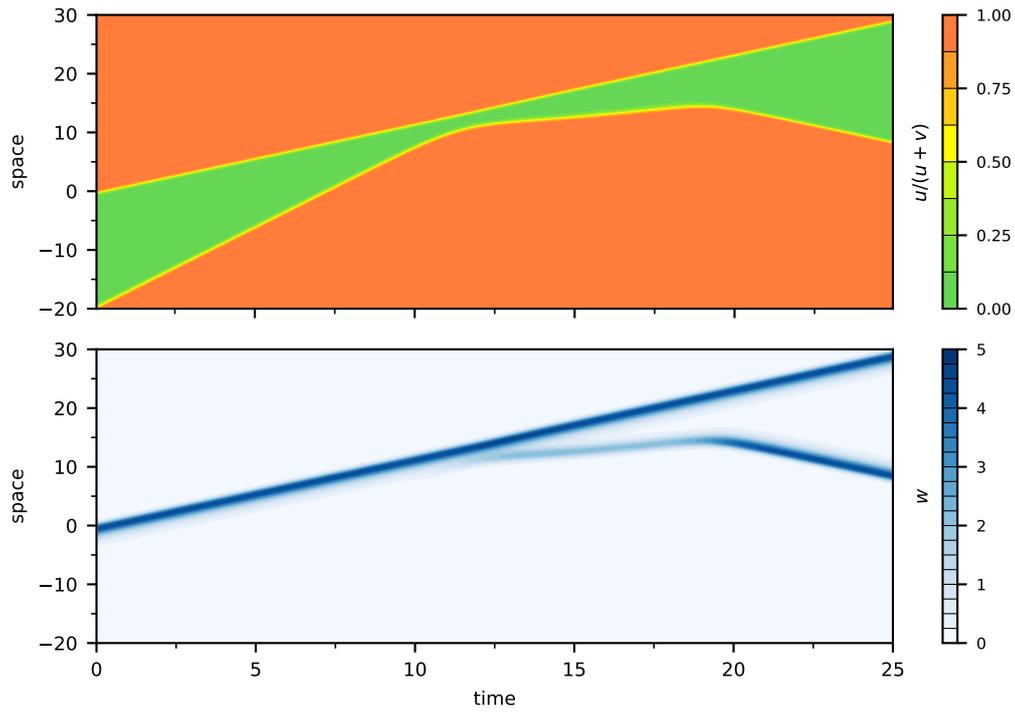

\else
\begin{figure}[p]
\fi
    \centering
    \includegraphics[width=0.87\textwidth]{{{tw_interaction__r3=26.8229__spacetime}}}
    \caption{Interaction of the trivial and non-trivial travelling waves for $r_3=26.8229$, resulting in their reflection.
             The unstable breathing wave exists for this value of $r_3$.}
    \label{fig:separator:reflection}
\ifchaos
\end{figure*}
\else
\end{figure}
\fi
\ifchaos
\begin{figure*}
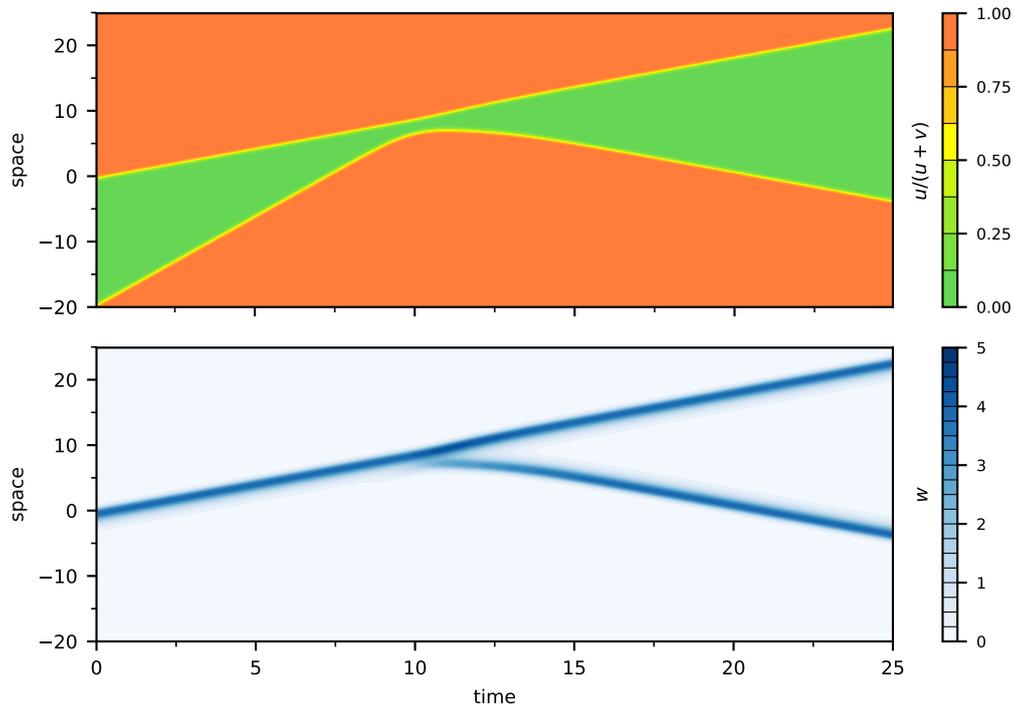

\else
\begin{figure}[p]
\fi
    \centering
    \includegraphics[width=0.87\textwidth]{{{tw_interaction__r3=26.75__spacetime}}}
    \caption{Interaction of the trivial and non-trivial travelling waves for $r_3=26.75$ (same simulation of Figure~\ref{fig:reflection}), resulting in their reflection.
             No breathing wave exists for this value of $r_3$.}
    \label{fig:separator:when_there_is_no_separator}
\ifchaos
\end{figure*}
\else
\end{figure}
\fi

Outcomes of pulse destabilization and travelling wave interaction are similar.
In fact, when the trivial wave approaches the slower non-trivial wave
their combined profile becomes pulse-like in appearance,
i.e., a sort of perturbation of the homoclinic wave.
As a consequence, the result of the interaction is determined by the fate of the perturbed pulse.
We checked this hypothesis by numerical simulation,
finding it true outside the region of existence of the unstable breathing wave ($r_{BF} < r_3 < r_{HC}$).
Examples of wave reflection for $r_3 < r_{BF}$ and wave merging for $r_3 > r_H$
have already been shown in Figures~\ref{fig:reflection} and \ref{fig:merging} respectively.
In Figure~\ref{fig:merging_to_breathing} we show an example
in which the colliding waves merge into the stable breathing travelling wave for $r_3 = 26.823 \in \left( r_{HC}, r_{H} \right)$
(the corresponding animation can be found at \intextcite{movieTW1d}).

In the region $r_{BF} < r_3 < r_{HC}$
a perturbed homoclinic wave becomes a breathing wave,
but the interaction of the trivial and non-trivial waves results in reflection,
as can be seen in Figure~\ref{fig:separator:reflection} for $r_3 = 26.8229$
(the corresponding animation can be found at \intextcite{movieTW1d}).
We believe this discrepancy is due to the existence of the unstable branch of the breathing travelling wave,
which acts as a separator between the stable breathing wave and the reflection-like behaviour.
An indication of this fact can be seen by looking closely at Figure~\ref{fig:separator:reflection}.
After the waves collide,
the back front does not immediately change direction of propagation,
but slows down first, resembling an unstable non-trivial wave.
This is the same behaviour we observed in the expansion phase of the unstable breathing wave
of Figure~\ref{fig:breathing_spacetime} (third row).
While in the unstable breathing wave
the exotic species $w$ present near the back front decays eventually
and two fronts approach again,
in the case of interacting waves
the back peak of the density $w$ grows to the full size observed in a stable non-trivial wave
and the back front starts to move backwards.
As a comparison, in Figure~\ref{fig:separator:when_there_is_no_separator}
we have plotted the space-time profile of the reflection of travelling waves
in the case $r_3 = 26.75$, when there is no unstable breathing wave.
In such a case, after collision the back front changes direction of propagation immediately,
skipping the intermediate phase in which the behaviour is similar to that of the unstable breathing wave.


\section{Interaction of stable trivial and non-trivial planar waves in two dimensions}
In order to explain the origin of the complex spatio-temporal pattern shown in Figure~\ref{fig:r_3_mid},
in this section we study the interaction of the planar extensions
of the trivial and non-trivial stable travelling waves introduced in Section~\ref{sec:1d_tw}.
As we have already observed, these two planar waves are stable in two spatial dimensions
for the parameter values satisfying \eqref{eq:parameters}.

As a simplified initial condition,
in a square domain $\Omega$
we place the non-trivial wave in the upper half-region
and the trivial wave in the lower half-region,
as shown in Figure~\ref{fig:2d_interaction_ic} (left panel).
The planar front velocities are respectively $c_{uvw}$ and $c_{uv}$,
where $c_{uvw} < 0$ varies with $r_3$
and $c_{uv} \approx 2.575 > 0$ is independent of $r_3$ instead.
Then, the non-trivial planar wave moves to the left in the upper part,
while the trivial wave moves to the right in the lower part.
This two-dimensional initial condition for the trivial and non-trivial fronts
is equivalent to the one shown in \cite[Figures~9 and 10]{coullet1992},
where the interaction of two Bloch fronts,
resulting in the appearance of a spiral pattern,
was considered in a system of forced coupled oscillators.

\ifchaos
\begin{figure*}
\else
\begin{figure}
\fi
    \centering
    \includegraphics[height=4cm]{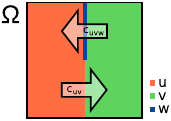}
    \qquad\qquad
    \includegraphics[height=4cm]{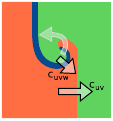}
    \caption{Left panel: Initial conditions for the interaction of non-trivial (top half-region) and trivial (bottom half-region) waves in a square domain $\Omega$.
             Right panel: Tip formation and movement when $\lvert c_{uvw} \rvert < \lvert c_{uv} \rvert$.
             Arrows outlined in black denote the front velocities, while the arrow with white outline shows the tip movement,
             which leads the tip front to collide with the tip back.}
    \label{fig:2d_interaction_ic}
\ifchaos
\end{figure*}
\else
\end{figure}
\fi

In order to similarly observe in our system
the appearance of a rotating tip at the point of contact of the two planar waves,
we will restrict ourselves to values of $r_3$ for which $\lvert c_{uvw} \rvert < \lvert c_{uv} \rvert$.
Then, the slower non-trivial wave makes up the back of this tip
while the faster trivial wave makes up its front,
as shown in Figure~\ref{fig:2d_interaction_ic} (right panel).
As a consequence, the front of the tip will eventually collide with its back.
From our observations in one spatial dimension presented in Section~\ref{sec:1d_tw_interaction},
we expect the results of this collision to depend on the relative velocity of the interacting fronts.
While in one spatial dimension
the relative velocity of the fronts
depends only on $r_3$,
in two (and above) dimensions
it depends also on the curvature of the fronts,
so that interacting fronts do not always behave similarly even if $r_3$ is the same.
We have already observed this phenomenon in our previous study of two-dimensional travelling waves of the three-species competition-diffusion system \cite{contento}.

We also remark that,
in order to keep computation time reasonable,
the spatial discretization used in our simulations
is rougher in two dimensions than in one dimension.
In particular, we have chosen a regular rectangular grid with step size $\Delta x = 0.1$.
Due to the different spatial resolution,
the exact values of $r_3$ at which the transitions between the different regimes occur will be different.
However, as far as we can tell, the qualitative behaviour does not change by further reducing $\Delta x$.
Moreover, if $\Delta x$ is too large (e.g., $\Delta x = 0.2$),
the anisotropy in the discretization
between directions parallel and diagonal to the axes will lead to a strong anisotropy in the front velocities.
This anisotropy can generate artefacts that destabilize the fronts.
In our future work, by applying a more sophisticated numerical method than finite differences,
we intend to simulate the two dimensional behaviour more precisely.
In particular, since the densities are essentially constants away from the moving interfaces,
adaptive mesh refinement seems to be a compelling option,
possibly using a posteriori error estimates
such as those recently proposed in \intextcite{morozovFEM}.

We have plotted the results of the numerical simulations
with the initial conditions of Figure~\ref{fig:2d_interaction_ic} (left panel)
in Figures~\ref{fig:2d_interaction:spiral}--\ref{fig:2d_interaction:complex_chaotic}.
The corresponding animations can be found at \intextcite{movieTW2d}.
First, suppose that $r_3$ lies in the region
where there exists a stable homoclinic travelling wave.
Then, we can expect the front of the tip to merge with its back when they collide.
As can be seen in Figure~\ref{fig:2d_interaction:spiral} for the example value $r_3 = 26.9$,
this leads to the formation
of a steadily rotating spiral
similar to those observed in excitable media,
such as the well known Belousov-Zhabotinsky chemical reaction \cite{youtubeBZ}.
Indeed, if we consider an initial condition
equal to $(r_1, 0, 0)$ in the upper half-region
and equal to the planar extension of the homoclinic wave in the lower part,
the end result is still a stable spiral.
Such initial conditions correspond
to ``cutting'' a front with an obstacle in the Belousov-Zhabotinsky reaction,
which is one of the standard methods used to obtain a spiral core.
Stable regular spiral waves also appear in systems with NIB bifurcations.
The case of forced coupled oscillators was studied in \cite{coullet1992} for the same initial conditions as ours,
while kinematic equations for the front motion in FitzHugh-Nagumo-like systems
were obtained in \cite{hagberg1997, hagberg1998} by singular perturbation in the vicinity of the bifurcation
and also showed the appearance of spiral patterns.

\ifchaos
\begin{figure*}
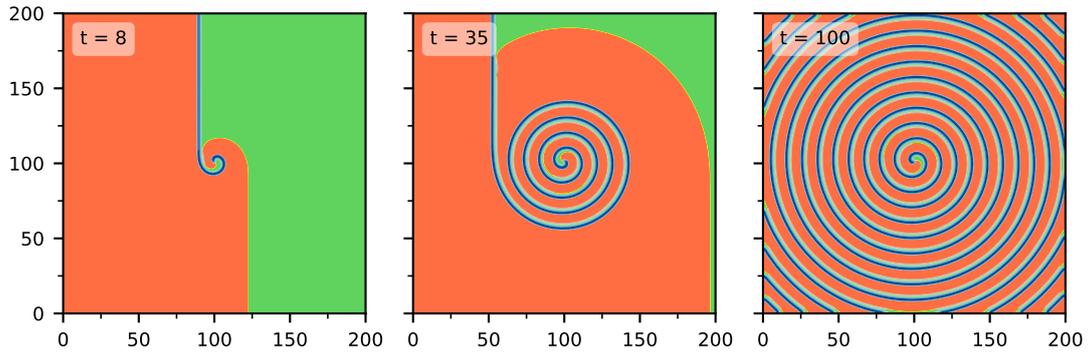

\else
\begin{figure}[p]
\fi
    \centering
    \includegraphics[width=0.93\textwidth]{{{tw_2d_interaction__r3=26.9000_dx=0.1}}}
    \caption{In the case $r_3=26.9$,
             initial conditions as in Figure~\ref{fig:2d_interaction_ic} (left panel)
             lead to a stable regular spiral.}
    \label{fig:2d_interaction:spiral}
\ifchaos
\end{figure*}
\else
\end{figure}
\fi

When $r_3$ decreases,
the homoclinic wave destabilizes by Hopf bifurcation
and a stable breathing wave appears.
The front of the tip still merges with its back
but the combined pulse breathes.
The final result is a spiral whose arm's width is oscillating.
When the oscillation is sufficiently small,
the coils of the spiral never touch and
this breathing spiral appears to be stable.
An example for $r_3 = 26.83$ is shown in Figure~\ref{fig:2d_interaction:breathing_spiral}.
Note that the arm width oscillates more strongly far away from the core.
We believe this is due to the fact that the high curvature of the trivial and non-trivial fronts near the core
decreases their relative velocity,
promoting their merging into a single pulse.
Then, the homoclinic pulse that makes up the spiral arm is more stable near the core
and the further it travels the more it oscillates.

It looks like that the regular spiral has destabilized by a Hopf bifurcation
generating a periodic pattern as a result,
behaving exactly as its one-dimensional counterpart, the travelling pulse
(more detailed computations are needed to confirm this intuition).
Breathing spirals of this kind are not common in the existing literature.
For example, they have been observed in the CDIMA chemical reaction,
but only under external periodic forcing \cite{berenstein2008, ghosh2011}.
On the other hand, breathing spirals have been shown to exist in absence of external forcing
in the Aliev–Panfilov model for the excitation dynamics of the cardiac tissue,
but only in annulus domains \cite{sakaguchi2010}.

\ifchaos
\begin{figure*}
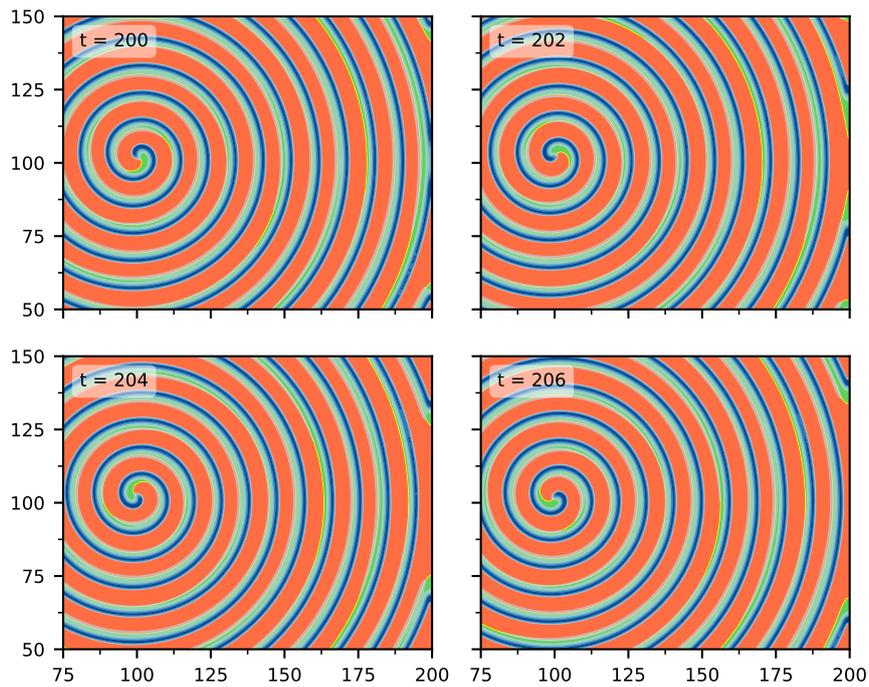

\else
\begin{figure}[p]
\fi
    \centering
    \includegraphics{{{tw_2d_interaction__r3=26.83}}}
    \caption{In the case $r_3=26.83$,
             initial conditions as in Figure~\ref{fig:2d_interaction_ic} (left panel)
             lead to a stable breathing spiral.
             The initial phase in which the spiral forms is similar to Figure~\ref{fig:2d_interaction:spiral}.
             Here we have plotted an enlarged region in order to better see the arm width variations.
             We encourage the reader to zoom in in the electronic version.}
    \label{fig:2d_interaction:breathing_spiral}
\ifchaos
\end{figure*}
\else
\end{figure}
\fi

If $r_3$ decreases further, the oscillations become larger
and eventually the homoclinic front splits into two non-trivial fronts.
Then, the coils of the spiral collide and annihilate,
leaving gaps in the spiral arm.
This leads eventually to the formation of new spiral cores.
An example of this process for $r_3=26.8175$ is shown in Figure~\ref{fig:2d_interaction:complex_periodic_large_cores}.
When the spiral is small, the stabilizing effect of curvature prevents it from breaking.
However, the spiral arm already starts to breath, as can be seen in Figure~\ref{fig:2d_interaction:complex_periodic_large_cores} (first panel).
When the spiral arm travels further enough from the core,
the oscillations become so large that the back front of a coil collides with the leading front of the previous coil,
resulting in their annihilation and in a gap in the spiral arm,
as can be seen in Figure~\ref{fig:2d_interaction:complex_periodic_large_cores} (panels 2--4).
The gap leads to the formation of new spiral cores and the same process repeats itself until a pattern formed by many spirals is obtained,
as can be seen in Figure~\ref{fig:2d_interaction:complex_periodic_large_cores} (panels 5--7).
The spiral cores are quite stable and do not display meandering.
For this reason, after an initial transient phase
the number and position of the spiral cores appear to become fixed
and the resulting pattern repeats itself periodically in time,
as can be seen by comparing the eight and ninth panels of Figure~\ref{fig:2d_interaction:complex_periodic_large_cores}.

There are two points we want to highlight.
First, the final pattern of the simulation in Figure~\ref{fig:2d_interaction:complex_periodic_large_cores}
consists in a large central spiral surrounded by a large number of smaller spirals.
Since a spiral can only grow as long as it does not destabilize or collide with another one,
only the initial spiral is able to reach its maximum allowed size.
The spiral cores generated after its breakdown must compete for space with each other,
preventing them from growing as much.
Thus, the particular distribution of spiral sizes observed
is not general but just a consequence of our choice of the initial conditions.
The only quantity independent of the initial conditions (but dependent on $r_3$)
is the maximum size that can be reached by a single spiral.
Secondly, spiral breakup occurs only far from the spiral core.
It follows that, if the same initial conditions for the same value of $r_3$
were to be considered in a smaller domain,
the resulting pattern would be a breathing spiral
such as the one shown in Figure~\ref{fig:2d_interaction:breathing_spiral}.
This raises the question of whether breathing spirals can only appear in bounded domains.
While rigorously proving the existence of breathing spirals in the whole $\mathbb{R}^2$ is an open problem,
we expect them to exist as long as $r_3$ is chosen sufficiently near $r_H$,
the Hopf bifurcation point of the homoclinic wave.

The mechanism leading to the generation of new spiral cores in Figure~\ref{fig:2d_interaction:complex_periodic_large_cores}
appears qualitatively similar to the spiral breakup
observed in FitzHugh-Nagumo-like systems
\cite{panfilov, hagberg1994patterns}.
In such cases the breakup of the spiral arm
is caused by the planar instability of one of its constituent fronts,
as for example it was shown
by singular limit analysis
for the case of Bloch fronts
in \cite{hagberg1997, hagberg1998}.
However, in our case both the trivial and non-trivial fronts, as well as the travelling pulse, are planarly stable
and thus we expect that breakup is caused by the instability of the breathing wave train associated to the spiral.
In addition, the destabilization process may be influenced
by the phase of the breathing oscillation
which changes along the spiral arm.
More detailed numerical investigation are needed to shed light on the precise mechanism.

Another difference from the above-cited examples
is that in our case the final pattern observed
is not chaotic spiral turbulence,
but a spatially-complex and time-periodic multi-core pattern.
We believe that even in FitzHugh-Nagumo-like systems,
by choosing initial conditions seeded with additional spiral cores
and setting the parameters so that no planar instability occurs,
the appearance of a qualitatively similar pattern is possible.
Indeed, this has been reported in \cite[p.~125]{coullet1992}
for a system of forced coupled oscillators.
However, we remark that in our case such complex pattern
can be generated from a very simple one-core initial condition
thanks to spiral breakup, which limits the size a single spiral can achieve.

\ifchaos
\begin{figure*}
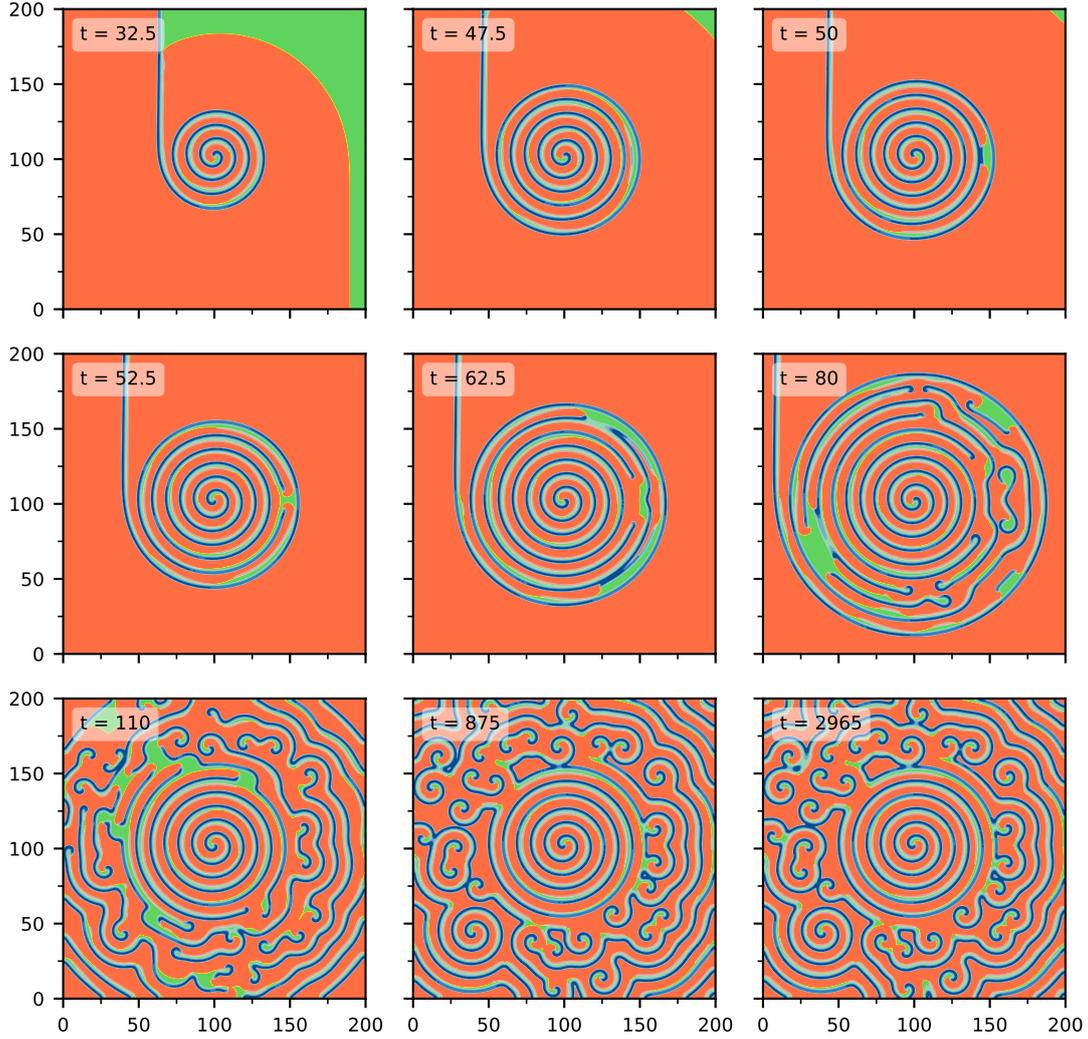

\else
\begin{figure}[t]
\fi
    \centering
    \includegraphics[width=0.93\textwidth]{{{tw_2d_interaction__r3=26.8175}}}
    \caption{In the case $r_3=26.8175$,
             initial conditions as in Figure~\ref{fig:2d_interaction_ic} (left panel)
             lead to a complex periodic pattern characterized by many spiral cores.}
    \label{fig:2d_interaction:complex_periodic_large_cores}
\ifchaos
\end{figure*}
\else
\end{figure}
\fi

As $r_3$ is reduced, the breakup starts to occur nearer to the spiral core.
As a consequence, the central spiral grows less before destabilizing
and the number of new spiral cores increases.
The pattern still behaves periodically in the long run,
but the initial transient phase is longer.
Moreover, it becomes possible that
the initial spiral
is no longer included in the final pattern
and disappears by interaction with the nearby spiral cores,
as can be seen in Figure~\ref{fig:2d_interaction:complex_periodic_small_cores} for $r_3=26.8$.
It should be noted that, although $r_3=26.8$ already lies inside the parameter region for which reflection occurs in one spatial dimension,
due to the effect of curvature many front interactions do not result in complete reflection,
but display breathing behaviour.

\ifchaos
\begin{figure*}
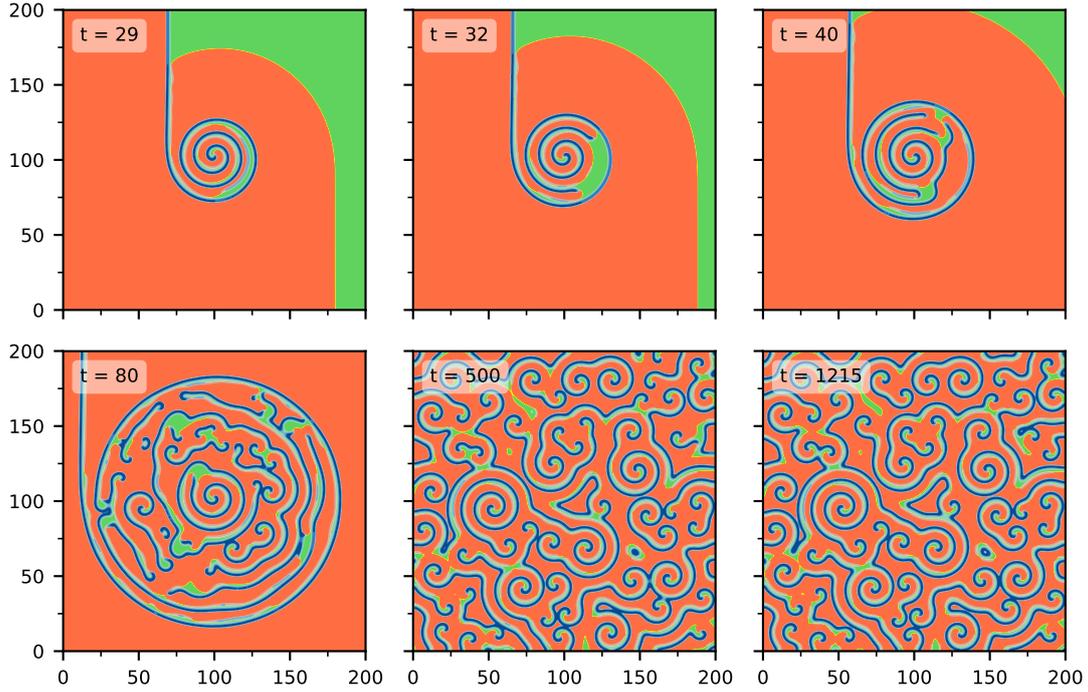

\else
\begin{figure}[p]
\fi
    \centering
    \includegraphics[width=0.93\textwidth]{{{tw_2d_interaction__r3=26.8000_dx=0.10}}}
    \caption{In the case $r_3=26.8$,
             initial conditions as in Figure~\ref{fig:2d_interaction_ic} (left panel)
             lead to a complex periodic pattern characterized by many spiral cores,
             as it was the case in Figure~\ref{fig:2d_interaction:complex_periodic_large_cores}.
             However, the initial spiral no longer survives in the long run.}
    \label{fig:2d_interaction:complex_periodic_small_cores}
\ifchaos
\end{figure*}
\else
\end{figure}
\fi
\ifchaos
\begin{figure*}
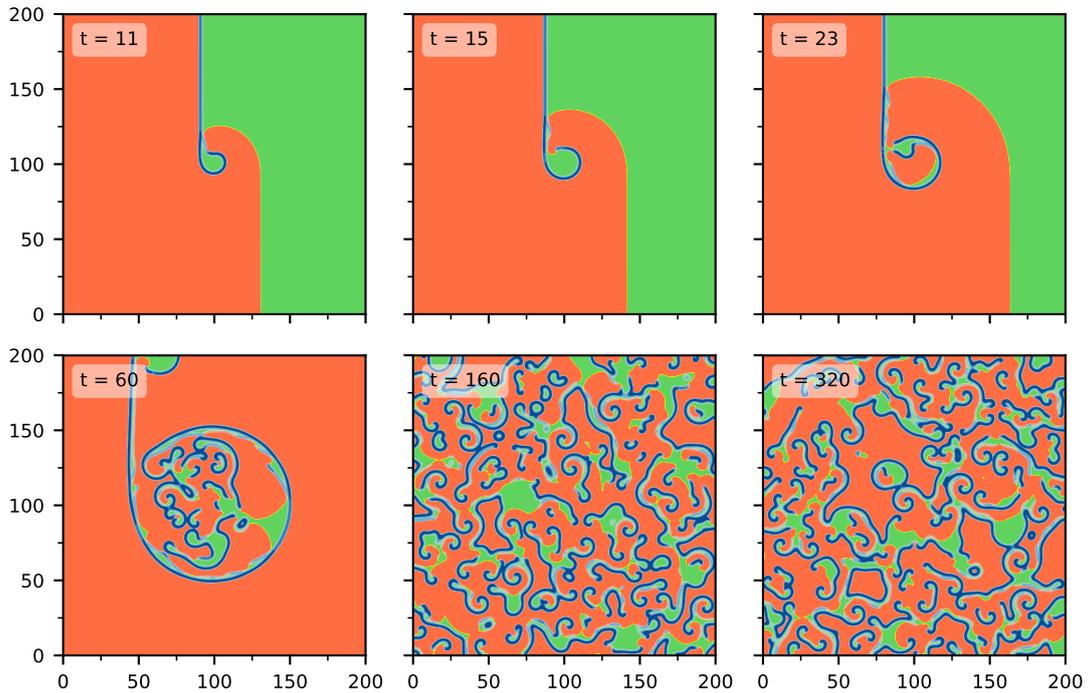

\else
\begin{figure}[p]
\fi
    \centering
    \includegraphics[width=0.93\textwidth]{{{tw_2d_interaction__r3=26.7500_dx=0.10}}}
    \caption{In the case $r_3=26.75$,
             initial conditions as in Figure~\ref{fig:2d_interaction_ic} (left panel)
             lead to a complex non-periodic pattern characterized by the continuous birth and destruction of small spiral cores.}
    \label{fig:2d_interaction:complex_chaotic}
\ifchaos
\end{figure*}
\else
\end{figure}
\fi

A further reduction of the parameter $r_3$
results in the central main spiral no longer being formed.
New spiral cores are still generated by the incomplete reflection mechanism,
although no spiral grows around them,
as can be seen in Figure~\ref{fig:2d_interaction:complex_chaotic} (panels 1--4) for $r_3=26.75$.
Since front interactions result in the complete reflection of fronts more often than before,
no spiral core is able to survive for a long time.
Old cores are continuously destroyed and new ones generated,
leading to a very complex spatio-temporal pattern
which is no longer periodic but chaotic in appearance,
as can be seen in Figure~\ref{fig:2d_interaction:complex_chaotic} (panels 5--6).
This is the same type of pattern we observed in Figure~\ref{fig:r_3_mid} at the beginning of this paper.

\section{Concluding remarks}
In this paper we studied
the invasion by an exotic species $w$
of an ecosystem inhabited by two native species $u$ and $v$.
The invader was supposed to be relatively weak
compared to the native species,
since it evolved in a different environment.
By using a three-species competition-diffusion system
and taking $r_3$, the intrinsic growth rate of $w$, as a free parameter,
we investigated whether competitive exclusion or competitor-mediated coexistence occurs.
Sensitively depending on the value of $r_3$,
we found that in two spatial dimensions
the system generates very complex spatio-temporal patterns
which allow the three species to coexist dynamically.

The main purpose of this paper was to reveal the mechanism behind the occurrence of such complex patterns.
From a visual inspection of the simulation results,
it is clear that the patterns are generated
by the interaction of two different kinds of moving fronts.
For this reason, we started by studying
the one-dimensional travelling waves of the system,
finding that for intermediate values of $r_3$
there exist two types of stable waves.
When their interaction is considered,
depending on their relative velocity $\Delta c$
different outcomes can be observed.
If $\Delta c$ is small in modulus,
the two waves merge into a single travelling pulse.
As $\Delta c$ becomes larger,
this pulse is destabilized through Hopf bifurcation
and a travelling pulse whose width oscillates (breathing wave) can be observed.
Then, colliding waves still merge, but become a breathing wave instead.
Finally, when $\Delta c$ is quite large the breathing wave no longer exists
and colliding waves are reflected.

After that,
we considered the two-dimensional case again,
studying the interaction of the planar extensions of the two stable travelling waves,
which for equal diffusion coefficients are also stable.
Using the insight obtained from the one-dimensional case,
we were able to see more clearly that a similar transition occurs in two spatial dimensions,
explaining the onset of the complex patterns we observed.
First, for $\Delta c$ small in modulus, a regular and stable spiral exists
in association to the one-dimensional stable pulse.
As $\Delta c$ becomes larger, the pulse width starts to oscillate:
the regular spiral becomes unstable,
but we still have a stable breathing spiral whose arm width oscillates.
A further increase of $\Delta c$ leads to larger oscillations
which eventually cause the spiral to break up,
generating additional spiral cores.
The resulting pattern appears to be periodic,
with the maximum possible spiral size decreasing as $\Delta c$ becomes larger.
At a certain point,
reflection of fronts becomes more common than oscillating behaviour.
As a consequence, the spiral cores are no longer persistent, but are dynamically destroyed and generated,
resulting in an extremely complex non-periodic pattern.

The main ingredients for the appearance of the complex patterns are thus two,
the existence of spiral cores and the reflection/breathing dynamics.
The spiral cores allow the species to coexist locally,
thus preventing competitive exclusion, which would result in the absence of any pattern.
Reflection-like front interaction prevents the formation of spirals and eventually destroys the spiral cores,
while breathing-like interaction leads to the formation of new spiral cores.
Thanks to the effect of front curvature
both reflection and breathing behaviour can coexist
for a range of values of $r_3$,
in which case
we can observe a complex and chaotic pattern
characterized by the dynamic generation and destruction of new spiral cores.

The same mechanism of spiral generation from a stable travelling pulse
is observed in many FitzHugh-Nagumo-like reaction-diffusion systems.
Additionally, in such systems spiral breakup and spiral turbulence are often observed
when the pulse is no longer planarly stable \cite{coullet1992, hagberg1994patterns, panfilov},
leading to patterns qualitatively similar to that in Figure~\ref{fig:r_3_mid}.
In particular, our system shares many similarities
with the cases in which two stable Bloch fronts
are generated from a nonequilibrium Ising-Bloch (NIB) bifurcation.
These two Bloch fronts connect the same asymptotic states but travel in opposite directions,
interact in complex ways and may exhibit breathing behaviour \cite{hagberg1994},
making them very similar to the trivial and non-trivial fronts studied in this paper.

We conclude by summarizing what we believe are the most novel points introduced in this work.
\begin{itemize}
\item In the case of a system with a NIB bifurcation
the interaction of two stable non-trivial (Bloch) fronts
bifurcating from the same trivial (Ising) front
is considered with similar results.
However, in our case the interacting fronts
are a stable non-trivial front and the stable trivial front,
since the second non-trivial front is unstable.
This is not the same as considering the imperfect pitchfork bifurcation obtained by perturbing the NIB (pitchfork) bifurcation \cite{meron},
since the non-trivial front is still generated from the trivial front by a drift bifurcation.
It may be possible to find a parameter set for which even in our case the drift bifurcation is a pitchfork (and thus a NIB) bifurcation,
but in that case we expect one of the Bloch fronts to display negative density values for the species $w$ and thus be non-admissible.
\item We have shown how the existence of a breathing travelling pulse is very important
in order to explain the outcome of pattern formation.
In particular, the unstable breathing wave acts as a separator between reflection and merging dynamics,
fulfilling a similar role to the scattors in the Gray-Scott model \cite{nishiuraInteraction}.
Finally, the bifurcation structure of the unstable branch as the period tends to infinity is quite interesting from a mathematical point of view
and should be object of a deeper investigation.
\item While the exact mechanism leading to spiral breakup
will require further investigations to be precisely determined,
we believe it to be linked to the destabilization of the time-periodic wave train associated to the breathing spiral.
The varying phase of the breathing oscillation along the spiral arm may also be an influencing factor.
Planar instability does not seem to play a role,
since the trivial and non-trivial fronts, together with the travelling/breathing pulse,
appear to be planarly stable
when the diffusion coefficients of all three species are equal.
This makes our case different from spiral breakup due to lateral instability,
such as that reported in \cite{hagberg1994patterns} and studied by singular perturbation in \cite{hagberg1997, hagberg1998}.
\item By the time spiral break-up occurs
the regular spiral has already been destabilized by Hopf bifurcation,
generating a stable breathing spiral.
Such breathing pattern, except for the cases of forced systems \cite{berenstein2008, ghosh2011} or non-convex domains \cite{sakaguchi2010},
is still not widely documented.
\item For a given value of the free parameter $r_3$,
the observed two-dimensional patterns are not completely determined
by the one-dimensional travelling wave behaviour
because curvature plays an important role.
Higher front curvature promotes merging of the fronts,
allowing both reflection and breathing dynamics to coexist,
enlarging the range of parameters for which coexistence is possible
compared to what could be expected by just considering the one-dimensional case.
Moreover, this mechanism prevents any spiral to grow too large,
allowing the generation of a periodic many-core pattern
even in the presence of a very simple initial condition.
\end{itemize}

\ifchaos
\begin{acknowledgments}
\else
\section*{Acknowledgements}
\fi
LC has been supported by the Meiji Institute for Advanced Study of Mathematical Sciences
and by JSPS KAKENHI Grant-in-Aid for Research Activity Start-up JP16H07254.
MM has been partially supported by JSPS KAKENHI Grant No.~15K13462.
\ifchaos
\end{acknowledgments}
\fi


\ifchaos
\else
\ifphysicaD
\bibliography{cm2018}
\else
\printbibliography{}
\fi
\fi

\end{document}